\pdfoutput=1
\documentclass[11pt]{article}
\usepackage{amsfonts}
\usepackage{amsmath}
\usepackage{hyperref}
\usepackage{amssymb}
\usepackage[english]{babel}
\usepackage{graphicx}
\usepackage{simplewick}
\usepackage{color}
\usepackage{soul}
\usepackage{caption}
\usepackage{subcaption}
\usepackage{booktabs}
\usepackage{cite}
\usepackage[utf8]{inputenc}

\def\fieldOne{\Phi^{(1)}}
\def\fieldTwo{\Phi^{(2)}}
\def\field{\Phi^{(i)}}
\def\cfieldOne{\tilde\Phi^{(1)}}
\def\cfieldTwo{\tilde\Phi^{(2)}}
\def\cfield{\tilde\Phi^{(i)}}

\def\ci{\text{i}}
\def\DM{\text{DM}}

\begin{document}
\allowdisplaybreaks

\begin{center}
{
\bf\Large
${\mathbb A}_4$ symmetry at colliders and in the universe
}
\\[8mm]
Ivo de Medeiros Varzielas$^{\star}$
\footnote{E-mail: \texttt{ivo.de@soton.ac.uk}}
,
Oliver Fischer$^{\dagger}$
\footnote{E-mail: \texttt{oliver.fischer@unibas.ch}}
and
Vinzenz Maurer$^{\dagger}$
\footnote{E-mail: \texttt{vinzenz.maurer@unibas.ch}}

\end{center}
\vspace*{0.50cm}
\centerline{$^{\star}$ \it
School of Physics and Astronomy, University of Southampton,}
\centerline{\it
Southampton, SO17 1BJ, U.K.}
\vspace*{0.2cm}
\centerline{$^{\dagger}$ \it
 Department of Physics, University of Basel,}
\centerline{\it
Klingelbergstr.~82, CH-4056 Basel, Switzerland}
\vspace*{1.20cm}
\begin{abstract}
\noindent
Two puzzling facts of our time are the observed patterns in the fermion masses and mixings and the existence of non-baryonic dark matter, which are both often associated with extensions of the Standard Model at higher energy scales.
In this paper, we consider a solution to these two problems with the flavour symmetry ${\mathbb A}_4\times {\mathbb Z}_2\times {\mathbb Z}_2^\prime$, in a model which has been shown before to explain large leptonic mixings with a specific texture.
The model contains 3 generations of $SU(2)_L$-doublet scalar fields, arranged as an ${\mathbb A}_4$-triplet, that spontaneously break the electroweak symmetry,
and a ``dark sector'' of ${\mathbb Z}_2$-odd fields, containing one Majorana neutrino and an ${\mathbb A}_4$-triplet $SU(2)_L$-doublet scalar field, the lightest of which provides a candidate for dark matter.

Concerning the ${\mathbb Z}_2$-even scalar fields, compared to the Standard Model, we predict additional fields with masses at the electroweak scale.
We therefore investigate present phenomenological constraints from lepton flavour violation experiments, obtaining a lower bound on the extra scalar masses of 140 GeV.
Furthermore we consider the oblique parameters, Higgs boson decay properties and possible flavour violating signals at the LHC.

Concerning the ``dark sector'', we study bounds from dark matter search experiments and identify the parameter space of the dark matter candidate that is compatible with the observed relic density.
We find two allowed mass ranges for the dark matter within which the experimental constraints can be accommodated: the low-mass range is from 47 GeV to 74 GeV and the high-mass range is from 600 GeV and 3.6 TeV.
\end{abstract}

\section{Introduction}

The Standard Model (SM) continues to be an extremely successful description of particle physics.
The scalar resonance that was observed by the ATLAS and CMS collaborations \cite{Aad:2012tfa,Chatrchyan:2012ufa}, with mass 125.7$\pm 0.4$ GeV \cite{Agashe:2014kda} 
is consistent with the SM Higgs boson. Nevertheless, there is still room for extensions of the scalar sector and
despite the discovery of a SM-like Higgs, the motivation for physics beyond the SM remains with the many unresolved puzzles that are not addressed by the SM. 
One of these puzzling facts is the existence of non-baryonic dark matter (DM), which is about five times more abundant than baryonic matter \cite{Ade:2013zuv}.
Up to now it is undetected by direct detection experiments and potential hints for DM signals in indirect detection experiments 
are not always conclusive due to astrophysical backgrounds, see e.g.\ refs.~\cite{Yuan:2014rca,Gomez-Vargas:2014gra,Calore:2015nua,O'Leary:2015gfa}.  
A new unbroken symmetry seems to be of fundamental importance to explain the DM properties: it prevents the DM particle from decaying into SM particles, and thus stabilises it while also explaining the suppressed interactions with the SM particles. 
A simplistic ``bottom-up'' approach to the DM problem is given by the class of one-particle extensions of the SM.
A more ``top-down'' approach often generates an entire ``dark sector'', as is the case e.g.\ for supersymmetry, extra dimensions and grand unified theories.

Another puzzling fact that remains unresolved within the SM is the observation of regularities in the fermion sector of the SM. 
The mere existence of three fermionic generations cannot be addressed or understood within the SM \footnote{An argument has been presented in \cite{vanderBij:2010nu} which connects the SM gauge group with the number of fermion generations.}, let alone the particular pattern of the fermion masses or their mixings.
A very peculiar feature is also found in the lepton sector, which shows at the same time tiny neutrino masses and large mixings between the generations.
This is a pattern that is different from what is observed for the quark sector.

An extension of the SM by one or more additional Higgs doublets \cite{Lee:1973iz} (see e.g.\ ref.~\cite{Branco:2011iw} for a review with the focus on two-Higgs-doublet models) can address both of the above mentioned shortcomings of the SM. One particularly intriguing possibility is to introduce the additional doublets as a multiplet of a flavour symmetry such as $\mathbb{A}_4$ \cite{Ma:2001dn, Lavoura:2007dw, Morisi:2009sc, Ma:2010gs}, see also ref.\ \cite{Hernandez:2013dta} and references therein.
In such models with a flavour-symmetric extension of the scalar sector 
the Higgs phenomenology leads to important direct constraints, while indirect constraints are given by fermion flavour violating processes or simply due to the requirement that the fermion masses and mixing are viable \cite{Toorop:2010ex, Toorop:2010kt, Felipe:2014zka}.
In this type of $\mathbb{A}_4$ model it is possible to account for an excess in $\mu \tau$ events that was found by the CMS collaboration \cite{CMS:2014hha, Khachatryan:2015kon}, which is interpreted as lepton-flavour-violating decays of the SM-like Higgs boson, see for instance ref.\ \cite{Campos:2014zaa,Sierra:2014nqa,Heeck:2014qea}

In this context, a very interesting ``top-down'' approach is given by the connection of the DM with the flavour sector where the DM candidate can be stabilised by the same symmetries that are introduced to explain patterns of leptonic mass and mixing, see e.g.\ refs.\ \cite{Hirsch:2010ru, Boucenna:2011tj, Ma:2012ez, Holthausen:2012wz, Ma:2014eka} for examples with $\mathbb{A}_4$ symmetries.
In flavour models with multiple scalar fields that have masses on the electroweak scale, i.e.\ ${\cal O}$(TeV), there is the promise of rich collider phenomenology that can potentially be explored with the next LHC run, at 13 TeV. In these models the DM candidate is typically part of a larger ``dark sector''.

In this paper we study a multi-Higgs and multi-component scalar DM model that entails a rich phenomenology, due in particular to the additional scalars having masses on the electroweak scale.
We show how existing measurements constrain the model and how it can be further tested in ongoing and future experiments, such as the next run of the LHC and dark matter search experiments.
The model is inspired by the $\mathbb{A}_4$ model in ref.~\cite{Hernandez:2013dta}, where the neutrino masses are generated radiatively and the observed large leptonic mixings can be reproduced.

This paper is structured in the following way: we review the relevant features of the model under consideration in section \ref{model}, in particular the scalar potential and the leptonic Yukawa couplings.
We investigate phenomenological constraints on the ${\mathbb Z}_2$-even scalar fields in section \ref{sec:H}, in particular from lepton flavour violating effects and Higgs decay properties.
In section \ref{sec:A} we consider the phenomenology of the ${\mathbb Z}_2$-odd scalar fields, where we in include constraints from astrophysics and direct searches at colliders. We conclude in section \ref{conclusion}. In appendices we include the details of $\mathbb{A}_4$ and also the ${\mathbb A}_4\times {\mathbb Z}_2 \times {\mathbb Z}_2^\prime$-invariant scalar potential used in this paper.

\section{The Model \label{model}}
The model considered in this paper consists in a multi-Higgs extension of the SM, with flavour symmetry $\mathbb{A}_{4} \otimes {\mathbb Z}_2 \otimes {\mathbb Z}^{\prime}_{2}$ \cite{Hernandez:2013dta}. The field content is summarised for convenience in table \ref{tab:assignment} with its symmetry assignments.

We embed the left-handed SM leptons $\ell_L$ into an $\mathbb{A}_4$-triplet and add one singlet Majorana neutrino $N_{R}$ that is odd under ${\mathbb Z}_2$.
The scalar sector contains two $SU(2)_{L}$-doublet $\mathbb{A}_4$-triplet fields $\Phi_{i}^{(1,2)}$, one of which also carries a ${\mathbb Z}_2$-charge,
and another $\mathbb{A}_4$-triplet field $\chi$, which is charged under the symmetry ${\mathbb Z}_2^\prime$. 
\begin{table}
  \centering
  \begin{tabular}{ccccccccc}
    \toprule
    & $\Phi^{(1)}$ & $\Phi^{(2)}$ & $\chi$ & $\ell_L$ & $e_R$ & $\mu_R$ & $\tau_R$ & $N_R$ \\
    \midrule
    $\mathbb{A}_{4}$ & $\mathbf{3}$ &$\mathbf{3}$ & $\mathbf{3}$ & $\mathbf{3}$ & $\mathbf{1}$  & $\mathbf{1'}$  & $\mathbf{1''}$  & $\mathbf{1}$ \\
    $SU(2)$ & $\mathbf{2}$  & $\mathbf{2}$  & $\mathbf{1}$  & $\mathbf{2}$  & $\mathbf{1}$  & $\mathbf{1}$  & $\mathbf{1}$ & $\mathbf{1}$ \\
    ${\mathbb Z}_{2}$ & $-1$ & $1$ & $1$ & $1$ & $1$ & $1$ & $1$ & $-1$ \\
    ${\mathbb Z}_{2}^\prime$ & $1$ & $1$ & $-1$ & $1$ & $1$ & $1$ & $1$ & $1$ \\
    $U(1)_{\textrm{Y}}$ & $1$ & $1$ & $0$ & $-1$ & $-2$ & $-2$ & $-2$ & $0$ \\  
    \bottomrule
  \end{tabular}
   \caption{Field content with charge assignment of the model. Note that $\chi$ is real.}
  \label{tab:assignment}
\end{table}
The Yukawa part of the model Lagrangian for the lepton sector takes the form
\begin{equation}
\begin{split}
\mathcal{L}_{Y} &= y_{e} \left( \overline{\ell}_{L}\Phi ^{\left( 2\right) }\right)_{\mathbf{\bf 1}} e_{R}
    + y_{\mu }\left( \overline{\ell}_{L}\Phi ^{\left( 2\right) }\right)_{\mathbf{1}^{\prime \prime }} \mu_{R}
    + y_{\tau }\left( \overline{\ell}_{L}\Phi^{\left( 2\right) }\right)_{\bf 1^{\prime }} \tau _{R} \\
&\quad + y_{\nu }\left( \overline{\ell}_{L}\widetilde{\Phi }^{(1)}\right)_{\mathbf{1}} N_{R}
    + M_{N}\overline{N}_{R} N_{R}^{c} + \text{h.c.} \,,
\end{split}
\label{LYlepton}
\end{equation}
with $\widetilde{\Phi}^{\left(k \right) }= \ci \sigma _{2}\left( \Phi ^{\left( k\right)}\right) ^{\ast }$ ($k=1,2$). The subscripts ${\bf 1, 1', 1''}$ denote the projection of the corresponding  $\mathbb{A}_{4}$ singlet in the product of the two triplets. For a brief summary of the $\mathbb{A}_4$ product rules, see Appendix \ref{A}. 
To be explicit we define the following decomposition of the scalar fields:
\begin{equation}\label{eq:fielddefinition}
    \Phi_{j}^{(1)} = \begin{pmatrix}
        \eta^\pm_j \\ 
        \frac{1}{\sqrt{2}}\left( \eta_{j} + \text{i}\, \tilde \eta_{j}\right)
    \end{pmatrix} \,, 
    \qquad
    \Phi_j^{(2)} = \begin{pmatrix}
        h^\pm_j \\ 
        \frac{1}{\sqrt{2}} \left(v_j + h_j +\text{i}\,\tilde h_j\right) 
    \end{pmatrix}\,,
    \qquad \text{with}\qquad j = 1,2,3\;.
\end{equation}
%
From the Yukawa terms in eq.~\eqref{LYlepton} the charged-lepton Yukawa matrices in the flavour eigenbasis $\hat Y_{i=1,2,3}$ for each component of  $\Phi^{(2)}_i$ follow:
\begin{equation}\label{eq:Yli}
    \hat Y_1 =
        \begin{pmatrix}
            y_e & y_\mu & y_\tau \\ 
            0 & 0 & 0 \\ 
            0 & 0 & 0  
        \end{pmatrix} \,, \enspace
    \hat Y_2 =
        \begin{pmatrix}
            0 & 0 & 0 \\ 
            y_e & \omega^2 y_\mu & \omega y_\tau \\ 
            0 & 0 & 0  
        \end{pmatrix} \,, \enspace
    \hat Y_3 =
        \begin{pmatrix} 
            0 & 0 & 0 \\ 
            0 & 0 & 0 \\ 
            y_e & \omega y_\mu & \omega^2 y_\tau 
        \end{pmatrix}
\end{equation}
Due to the ${\mathbb Z}_2$ symmetry, the field $N_R$ couples exclusively to the $\mathbb{A}_4$-triplet $SU(2)$-doublet $\Phi^{(1)}$. The Yukawa couplings for the neutrinos involving the components $\Phi^{(1)}_i$ and $\ell_{L,j}$ are given by $y_\nu \, \delta_{i j}$.

When the $\mathbb{A}_4$-triplet $SU(2)_L$-doublet $\Phi^{(2)}$ develops a non-zero vacuum expectation value (VEV), the Yukawa matrices in eq.~\eqref{eq:Yli} generate a mass matrix for the charged leptons.
Without breaking the ${\mathbb Z}_2$ symmetry, the masses for the left-handed neutrinos can be generated radiatively, via $\langle \Phi^{(2)}\rangle$ and $\langle\chi \rangle$.
The corresponding mass matrix of the neutrinos is sensitive to 
both VEVs, in contrast to the mass matrix of the charged leptons.
As was shown in ref.~\cite{Hernandez:2013dta}, in this case viable leptonic mixing is possible if the two VEVs $\langle \chi \rangle$ and $\langle \Phi^{(2)} \rangle$ have a specific {\it alignment} in $\mathbb{A}_4$-direction and the magnitudes of the VEVs exhibit a hierarchy
\begin{equation}\label{eq:approximation}
    \frac{\langle \Phi^{(2)} \rangle^2}{\langle \chi \rangle^2} =: r \ll 1\;.
\end{equation}
The most general scalar potential which is invariant under the $SU(2)_L\times \mathbb{A}_4 \times {\mathbb Z}_2 \times {\mathbb Z}'_2$-symmetry can be expressed as:
\begin{equation}\label{eq:potential}
    V = V_{\phi\phi}(\Phi^{(1)},\Phi^{(2)}) + V_{\chi \phi}(\chi,\Phi^{(1)},\Phi^{(2)}) + V_{\chi\chi}(\chi) \;.
\end{equation}
The three sub-potentials on the right-hand side of eq.~\eqref{eq:potential} are shown in full in Appendix \ref{B}. 
We find that for a suitable choice of the parameters, the sub-potential $V_{\chi\chi}$ can indeed cause the real scalar field $\chi$ to develop the VEV $\langle \chi \rangle$ and break the $\mathbb{A}_4$ symmetry.
The sub-potential $V_{\chi \phi}$ consists in the interaction terms between the $\chi$ fields and the $\Phi^{(i)},i=1,2$ fields. It also adds terms proportional to $\langle \chi \rangle$ to the mass matrices of the fields $\Phi^{(i)},\,i=1,2$.
The sub-potential $V_{\phi\phi}$ allows for electroweak symmetry breaking and adds interactions between the scalar fields $\Phi^{(1)}$ and $\Phi^{(2)}$. 
This sub-potential can lift the mass-degeneracy between the components of the three electroweak doublet generations contained in $\Phi^{(1)}$.

The special directions in $\mathbb{A}_4$-space of the VEVs $\langle \chi \rangle$ and $\langle \Phi^{(2)} \rangle$ are referred to as {\it VEV alignment}. The particular VEV alignment that was used in \cite{Hernandez:2013dta} in order to generate large leptonic mixing is
\begin{align}\label{eq:VEVs}
    \left\langle \Phi^{(1)} \right\rangle &= 0\,, \notag \\
    \left\langle \Phi^{(2)} \right\rangle &= \frac{v_\mathrm{EW}}{\sqrt{6}}
        \begin{pmatrix} \begin{matrix} 0 \\ 1 \end{matrix}& \begin{matrix} 0 \\ 1 \end{matrix} & \begin{matrix} 0 \\ 1 \end{matrix} \end{pmatrix}\,, \\
    \left\langle \chi \right\rangle &=  \frac{v_{\chi }}{\sqrt{2}}
        \begin{pmatrix} 1 & 0 & -1 \end{pmatrix} \,, \notag
\end{align}
with $v_\mathrm{EW} = 246.22$~GeV $\ll v_\chi$ being the electroweak vacuum expectation value, cf.\ eq.~\eqref{eq:fielddefinition}. The horizontal direction is the $\mathbb{A}_4$ direction where the vertical direction (displayed for  $\langle \Phi^{(2)} \rangle$) shows the  $SU(2)_L$ components.
Note that the $\Phi^{(1)}$ field does not develop a VEV, such that the ${\mathbb Z}_2$ symmetry remains unbroken.
For the potential in Appendix~\ref{B}, the direction of $\left\langle \chi \right\rangle$ is a stationary point 
but it is not a true minimum.
We discuss this issue in section \ref{sec:H} and in Appendix \ref{Ap:chis}.

With the VEV $\left\langle \Phi^{(2)} \right\rangle$ as in eq.~\eqref{eq:VEVs}, the mass and mixing matrix, respectively, for the charged leptons follow from  eq.~\eqref{eq:Yli}:
\begin{equation}\label{eq:chargedleptonmassmatrix}
    M^l = \frac{v_\mathrm{EW}}{\sqrt{6}}
        \begin{pmatrix} 
            y_e & y_\mu & y_\tau \\ 
            y_e & \omega^2 y_\mu & \omega y_\tau \\ 
            y_e & \omega y_\mu & \omega^2 y_\tau   
        \end{pmatrix}\,,
\qquad
V^l = \frac{1}{\sqrt{3}}
        \begin{pmatrix} 
            1 & 1 & 1 \\ 
            1 & \omega & \omega^2 \\ 
            1 & \omega^2 & \omega   
        \end{pmatrix}
 \,,
\end{equation}
such that $V^l M^l M^{l \dagger} V^{l \dagger}$ is diagonal and positive.
In turn, the structure of the light neutrino mass and mixing matrix, respectively, is dominated by $\left\langle \chi \right\rangle$ and takes the form
\begin{equation}
\def\arraystretch{1.3}
M^{\nu } \simeq 
    \begin{pmatrix}
        a \, e^{2i\psi } & 0 & a \\
        0 & b & 0 \\
        a & 0 & a \, e^{-2i\psi }%
    \end{pmatrix}%
    \,,
\qquad
V^ {\nu } \simeq
    \begin{pmatrix} 
        \frac{1}{\sqrt{2}} & 0 & \pm \frac{1}{\sqrt{2}}e^{2 i \psi} \\ 
        0 & 1 & 0 \\
        \mp \frac{1}{\sqrt{2}} e^{-2 i \psi} & 0 & \frac{1}{\sqrt{2}} 
    \end{pmatrix}
    \,
    P_\nu
\label{Mnu}
\def\arraystretch{1}
\end{equation}
where $a$ and $b$ are specific loop functions that depend on the Yukawa couplings, mass of the RH neutrino and the masses of the $\mathbb{Z}_2$-odd scalars, and 
$P_\nu$ is a diagonal phase matrix. $V_\nu$ and $V^l$ combine to a viable PMNS matrix as described in \cite{Hernandez:2013dta}.

We remark that in this model the assignment of quarks under $\mathbb{A}_4$ was not considered explicitly. 
In $\mathbb{A}_4$-symmetric models where the scalar $SU(2)_L$-doublets transform as triplets, the quark masses and mixings can be generated by embedding the three generations of up and down quarks into $\mathbb{A}_4$-triplet fields. Thereby either the left-handed $SU(2)_L$-doublet $Q_L$, the right-handed $SU(2)_L$-singlets $u_R$ and $d_R$, or both, can be defined to transform as $\mathbb{A}_4$-triplets. This type of assignment with quark generations embedded in triplets is usually considered in grand unified models featuring $\mathbb{A}_4$ to explain leptonic mixing \cite{deMedeirosVarzielas:2005qg} (see \cite{Bjorkeroth:2015ora} for a recent example), or otherwise to obtain relations between quarks and leptons \cite{Morisi:2011pt,King:2013hj}.

\section{${\mathbb Z}_2$-even Scalar Sector \label{sec:H}}
\subsection{Physical Field Content}
We observe that the VEV of the $\chi_i$ fields breaks the ${\mathbb Z}_2^\prime$ symmetry, such that they mix with the $\Phi^{(2)}$ fields.
The mass matrix of the full scalar sector is given by
\begin{equation}\label{eq:massmatrix}
    {\cal M}(X)_{i,j} = 
        \left\langle\frac{\partial^2 V}{\partial X_i\partial X_j}\right\rangle
\,, 
    \qquad 
    X = h, \tilde h, h^\pm, \eta, \tilde \eta, \eta^\pm,\chi \,,
\end{equation}
with the fields $h,\eta$ from the definition in eq.~\eqref{eq:fielddefinition} and where we suppressed the $\mathbb{A}_4$ indices. 
The charged fields $H^\pm_{i=1,2,3}$ are linear combinations of the charged $h_{j=1,2,3}^\pm$. 
The physical scalar fields $H_i,\,i=1,...,6$ and $\tilde H_{j=1,2,3}$ are linear combinations of the flavour eigenstates $h_k,\tilde h_k$ and $\chi_k$, with $k=1,2,3$. 
They are obtained from diagonalising the mass matrix in eq.~\eqref{eq:massmatrix}, which is block decoupled from the fields $\eta_{i=1,2,3}$ due to the unbroken ${\mathbb Z}_2$ symmetry.
In the limit $r=0$ the mixing between the $h$, the $\tilde h$ and the $\chi$ vanishes and the transformation between flavour basis and mass basis can be expressed as
\begin{align} 
    H = {\cal R}_H X &\qquad  X = \left(h_1, h_2, h_3, \chi_1, \chi_2, \chi_3 \right)^T\,,\notag \\
    \tilde H = {\cal R}_{\Phi_{(2)}} X & \qquad X = \left(\tilde h_1, \tilde h_2, \tilde h_3 \right)^T\,,\\
    H^\pm = {\cal R}_{\Phi_{(2)}} X & \qquad X = \left(h_1^\pm, h_2^\pm, h_3^\pm \right)^T \notag\,,
\end{align}
with the fields $h, \chi$ from the definition in eq.~\eqref{eq:fielddefinition}, the $6 \times 6$ matrix ${\cal R}_H$ being given by
\begin{equation}
\def\arraystretch{1.2}
    {\cal R}_H = \begin{pmatrix}{\cal R}_{\Phi_{(2)}} & 0  \\ 0 & {\cal R}_{\chi}  \end{pmatrix}\,,
\def\arraystretch{1}
\end{equation}
and the rotation matrices ${\cal R}_{\Phi_{(2)}}$ and ${\cal R}_{\chi}$
\begin{equation}
\def\arraystretch{1.5}
{\cal R}_{\Phi_{(2)}} = \begin{pmatrix} 
        \sqrt{\frac{1}{3}} & \sqrt{\frac{1}{3}} & \sqrt{\frac{1}{3}} \\ 
        -\sqrt{\frac{1}{6}} & \sqrt{\frac{2}{3}} & -\sqrt{\frac{1}{6}}\\
        \sqrt{\frac{1}{2}} & 0 & -\sqrt{\frac{1}{2}} 
    \end{pmatrix}\,,
\qquad
{\cal R}_{\chi} = \begin{pmatrix} 
        \sqrt{\frac{1}{2}} & 0 & \sqrt{\frac{1}{2}} \\ 
        0 & 1 & 0 \\
        -\sqrt{\frac{1}{2}} & 0 & \sqrt{\frac{1}{2}} 
    \end{pmatrix}\,.
\def\arraystretch{1}
\end{equation}
The mass matrix defined in eq.~\eqref{eq:massmatrix} contains some eigenvalues that vanish in the strict limit $r =0$, therefore we present the eigenvalues to leading order in $r$:
\begin{equation}\label{eq:higgsmasses}
\begin{split}
    m_{H_1}^2 &= 4 \left(\lambda_{1,(2)} + \lambda_{2,(2)} + \lambda_{3,(2)} - 
        \tfrac{\sigma _{1,(2)}{}^2}{4 d_1+d_2}\right) \, v_\mathrm{EW}^2 \,, \\
    m_{H_2}^2 &= 2 \, (\lambda _{4,(2)}+\lambda _{5,(2)}) \, v_\mathrm{EW}^2\,, \\
    m_{H_3}^2 &= 2 \, \sigma_{2,(2)} \, v_\chi^2\,, \\
    m_{H_4}^2 &= 4 \, (3 d_2 - 4 d_1) \, v_\chi^2\,, \\
    m_{H_5}^2 &= 2 \, (4 d_1 - 3 d_2) \, v_\chi^2\,, \\
    m_{H_6}^2 &= 4 \, (4 d_1 + d_2) \, v_\chi^2 \,. \\
\end{split}
\end{equation}
For the charged fields we get the following eigenvalues, using the same rotation matrix ${\cal R}_{\Phi_{(2)}}$,
\begin{equation}
\begin{split}
    m_{H_1^\pm}^2 &= 0 \,, \\
    m_{H_2^\pm}^2 &= (\lambda_{5,(2)} - 2 (\lambda_{2,(2)} + \lambda_{3,(2)})) \, v_\mathrm{EW}^2 \,, \\
    m_{H_3^\pm}^2 &= 2 \, \sigma_{2,(2)} \, v_\chi^2\,.
\end{split}
\end{equation}
Finally we find for the eigenvalues for the fields $\tilde H_{i=1,2,3}$, again using ${\cal R}_{\Phi_{(2)}}$:
\begin{equation}
\begin{split}
    m_{\tilde H_1}^2 & = 0 \,, \\
    m_{\tilde H_2}^2 & = 2 \, (\lambda _{5,(2)}-2 \lambda_{2,(2)}) \, v_\mathrm{EW}^2\,, \\
    m_{\tilde H_3}^2 & = 2 \, \sigma_{2,(2)} \, v_\chi^2\,.
\end{split}
\end{equation}
Among those, the neutral field $\tilde H_1$ and the charged field $H_1^\pm$ with mass eigenvalues zero can be identified as the Goldstone bosons. The field $H_1$ is the SM-like Higgs boson of our model, that we can identify with the scalar resonance at 125.7 $\pm$ 0.4 GeV \cite{Agashe:2014kda}.
It couples to the mass eigenstates of the leptons via the Yukawa matrices
\begin{equation}
Y_1 = \frac{1}{\sqrt{2}}\begin{pmatrix} 
        y_e & 0 & 0 \\
        0 & y_\mu & 0 \\
        0 & 0 & y_\tau
    \end{pmatrix}
    \times \left(1 - \left|\frac{\sigma_{1,(2)}}{4d_1+d_2}\right|^2 r\right)\,,
\end{equation}
which is proportional to the SM Higgs boson couplings with a reduction factor proportional to the small parameter $r$.

The second generation of ${\mathbb Z}_2$-even fields, namely $H_2$, $\tilde H_2$, $H_2^\pm$ have masses proportional to the VEV of $\Phi^{(2)}$ and thus also reside at the electroweak scale. They couple purely off-diagonally to the leptons, via the Yukawa matrix
\begin{equation}
Y_2 = \frac{1}{2}
    \begin{pmatrix} 
        0 & y_\mu \omega^2 & y_\tau \omega \\
        y_e \omega & 0 & y_\tau \omega^2 \\
        y_e \omega^2 & y_\mu \omega & 0
    \end{pmatrix}\,,
\label{eq:Y2}
\end{equation}
with $\omega = e^{\ci 2 \pi/3}$. The Yukawa matrices for the second generation of the ${\mathbb Z}_2$-even scalars are given by 
\begin{equation}\label{eq:2ndgenerationYukawa}
	( \tilde Y_2)_{\ell_L \ell'_R} = \ci \, \left(Y_2\right)_{\ell_L \ell'_R} \,, 
	\qquad 
	\left( Y_2^\pm \right)_{\ell_L \ell'_R} = \sqrt{2} \, \left(Y_2\right)_{\ell_L \ell'_R}\,,
\end{equation}
while the analogous Yukawa couplings for the third generation can be obtained using the substitutions $\omega\to -\ci \omega$ and $\omega^2 \to \ci \omega^2$ in eq.~\eqref{eq:Y2} and eq.~\eqref{eq:2ndgenerationYukawa}.
The fields $H_{4,5,6}$ do not couple to fermions at all, as they originate from $\chi$.
Note that beyond leading order $H_1$ mixes only with one state contained in $\chi$ (namely $H_6$) leading to the shown reduction factor compared to the SM and to the fact that all vanishing entries in Yukawa matrices also vanish for $r \ne 0$.

Notice that the squared masses of the $H_4$ and $H_5$ in eq.~\eqref{eq:higgsmasses} are not independent for $r=0$: $m_{H_5}^2 = -2\,m_{H_4}^2$. This is due to the VEV $\langle \chi \rangle$ being a saddle point solution rather than a minimum of the sub-potential $V_{\chi\chi}$.
One can impose the relation $d_1 = 3/4\, d_2$ to avoid a tachyonic mass. This restrictive region of parameter space then allows $\langle \chi \rangle$ to become a flat minimum where, however, even for non-zero $r$ one additional zero mass eigenvalue is present.
We present a simple solution to this issue in Appendix \ref{Ap:chis}, which does not affect the discussion of the phenomenology. For this reason, in what follows we shall consider the mass-scales of the $H_{i\geq 3}$ fields to be ${\cal O}(v_\chi)$.

\subsection{Phenomenology \label{sec:Hpheno}}
The ${\mathbb Z}_2$-even scalar fields of the considered model exhibit the peculiar feature of having two generations of fields that couple completely off-diagonally to the charged leptons. In this section we investigate the subsequent constraints from experimental bounds on lepton flavour violation, from electroweak precision data and from the decay properties of the 125.7 GeV SM-like Higgs boson.

\subsubsection{Lepton Flavour Violation}

\begin{figure}
        \centering
        \begin{subfigure}[b]{0.33\textwidth}
                \includegraphics[width=\textwidth]{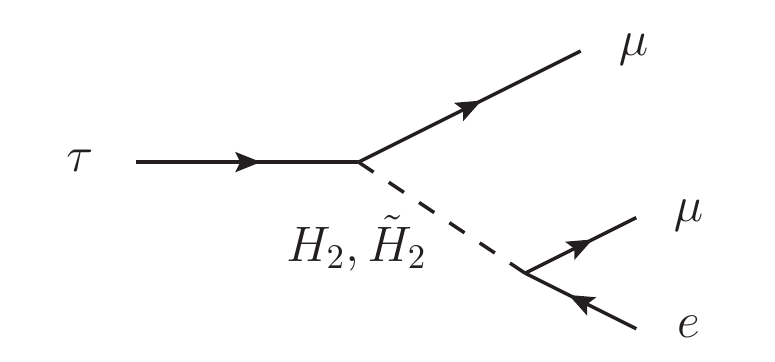}
                \caption{$\tau \to \mu \mu e$}
        \end{subfigure}
        \begin{subfigure}[b]{0.33\textwidth}
                \includegraphics[width=\textwidth]{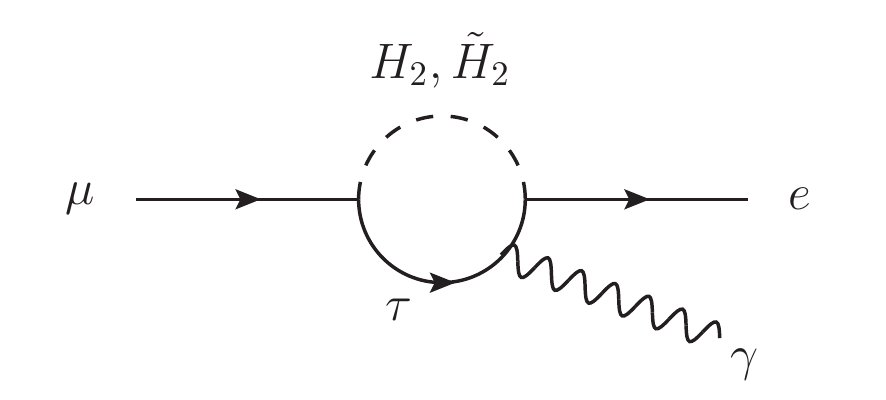}
                \caption{$\mu \to e \gamma$}
        \end{subfigure}
        \begin{subfigure}[b]{0.3\textwidth}
                \includegraphics[width=\textwidth]{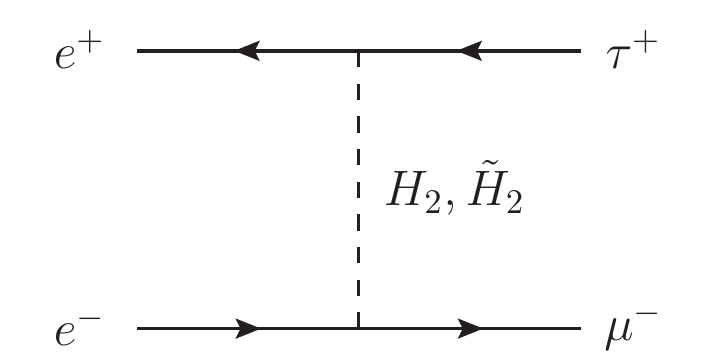}
                \caption{$e^+ e^- \to \tau^+ \mu^-$}
        \end{subfigure}
        \caption{Lepton flavour violating processes, mediated by the fields $H_2$ and $\tilde H_2$.}
        \label{fig:LFV}
\end{figure}

Lepton flavour violation (LFV) occurs at tree level in our model, due to the Yukawa couplings of the second generation ${\mathbb Z}_2$-even scalar fields ($H_2$, $\tilde{H}_2$ and $H_2^\pm$) -- the third generation in principle contributes similarly, but is assumed to decouple due to its mass of $\mathcal{O}(v_\chi)$ in the following.
It is interesting to note that the typically very constraining process $\mu \to e e e$ is not mediated at tree level by $H_2$ or $\tilde{H}_2$, since both fields have no flavour-conserving couplings to the leptons. Instead, the following processes are relevant:
\begin{equation}\label{eq:treelevelLFV}
    \tau^- \to e^- \mu^+ \mu^-\,,\; e^+ \mu^- \mu^-\,,\; \mu^- e^+ e^-\,, \text{ and } \mu^+ e^- e^-\,.
\end{equation}
For these processes we depict an example diagram in fig.~\ref{fig:LFV} (a), with the flavour-violating fields $\tilde H_2$ and $H_2$. 
The leading contributions to the lepton flavour violating tau decays of eq.~\eqref{eq:treelevelLFV} are given by
\begin{align}
    R_{\tau^- \to e^- \mu^+ \mu^-} &= \frac{
    2 \, m_{\mu}^4 \left(m_{\tilde{H}_2}^2 + m_{H_2}^2 \right)^2
    + m_{\tau}^2 \, m_{\mu}^2 \left(m_{H_2}^2 - m_{\tilde{H}_2}^2\right)^2
    }{128 \, m_{H_2}^4 m_{\tilde{H}_2}^4} \,,\\
    R_{\tau^- \to e^+ \mu^- \mu^-} &= \frac{
    m_{\mu}^4 \left(m_{H_2}^2 - m_{\tilde{H}_2}^2\right)^2 + 4 \, m_{\mu}^2 \, m_{\tau}^2 \left(m_{\tilde{H}_2}^2 + m_{H_2}^2\right)^2}{128 \, m_{H_2}^4 m_{\tilde{H}_2}^4} \,,\\
    R_{\tau^- \to \mu^- e^+ e^-} &= \frac{m_{\mu}^2 \, m_{\tau}^2 \left(m_{\tilde{H}_2}^2 + m_{H_2}^2\right)^2 }{64 \, m_{H_2}^4 m_{\tilde{H}_2}^4} \,,\\
    R_{\tau^- \to \mu^+ e^- e^-} &= \frac{
    m_{\mu}^2 \, m_{\tau}^2 \left(m_{H_2}^2 - m_{\tilde{H}_2}^2\right)^2}{128 \, m_{H_2}^4 m_{\tilde{H}_2}^4}\,,
\end{align}
where we defined $R_{\tau \to X}$ = BR($\tau^- \to X$)/BR($\tau^-\to e^- \bar{\nu}_e \nu_\tau$) and the lepton masses $m_\mu = 0.105$~GeV and $m_\tau = 1.77$~GeV. All formulae were obtained using FeynArts \cite{Hahn:2000kx} and FormCalc \cite{Hahn:1998yk,Hahn:2006qw}.
The process $\tau^- \to \mu^- e^+ e^-$ with the experimental upper bound being ${\cal O}(10^{-8})$ results in the very mild constraint on $m_{\tilde H_2},\,m_{H_2}^{} > 12$ GeV.
\medskip

Recent and very stringent constraints on rare charged lepton flavour violating decays come from the MEG collaboration \cite{Adam:2013mnn}. The contribution from the fields $H_2$ and $\tilde H_2$ to those rare decays is depicted in fig.\ \ref{fig:LFV} (b). With the definition $m_{\tilde{H}_2}^2 = M^2 + \Delta/2$, $m_{H_2}^2 = M^2 - \Delta/2$, the contribution can be approximated as (for not too large $\Delta$):
\begin{equation}\label{eq:mutoegamma}
\begin{split}
    \text{BR}(\mu \to e \gamma) &=
        \frac{\alpha \, m_{\tau}^4}{384 \pi M^8}
        \Big[
        4 M^4 - 60 \Delta M^2 + 227 \Delta^2 \\
        &
        \quad\quad + 24 \, \Delta \log\left(\frac{m_{\tau}}{M}\right) 
            \left(- 2 M^2 + 15 \Delta + 6 \Delta \log\left(\frac{m_{\tau}}{M}\right)\right) 
        \Big] \;.
\end{split}
\end{equation}
This was obtained using the formulae in ref.~\cite{Lavoura:2003xp}.

\begin{figure}
\centering
\includegraphics[width=0.40\textwidth]{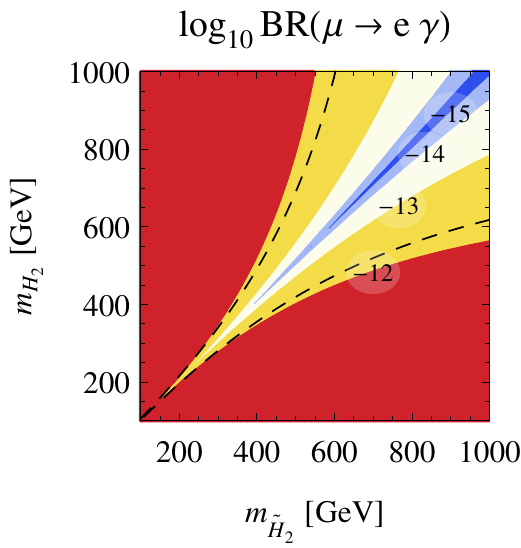}
\caption{BR$(\mu \to e\gamma)$ as mediated by the fields $H_2$ and $\tilde H_2$. The field $H_2^\pm$ contributes negligibly. The dashed line corresponds to the experimental constraint \mbox{BR$(\mu \to e\gamma) < 5.7 \cdot 10^{-13}$}~\cite{Agashe:2014kda}. }
\end{figure}

The charged field $H_2^\pm$ also contributes to the process, but according to the corresponding Yukawa matrix in eq.~\eqref{eq:2ndgenerationYukawa} its contributions to eq.~\eqref{eq:mutoegamma} are suppressed by the tiny ratios $m_e/m_\mu$ and $m_\mu/m_\tau$.
For the case $m_{\tilde H_2} \simeq m_{H_2}^{}$, the experimental bounds on the branching ratio of BR$(\mu \to e \gamma) < 5.7\times 10^{-13}$ then requires $m_{H_2}>143$ GeV. 
We note that when the model saturates the MEG bound, i.e.\ produces a branching ratio of $\mu \to e\gamma$ comparable with the current sensitivity limit, the branching ratios for the similar processes in $\tau$ decays are given by
\begin{equation}\label{eq:taugammadecays}
    \text{BR}(\tau \to \mu \gamma) = 2.1\times 10^{-16} 
    \qquad \text{and} \qquad 
    \text{BR}(\tau \to e\gamma) = 1.5\times 10^{-18}\,.
\end{equation}
\medskip

Bounds on LFV have been established at LEP-II, for instance by the OPAL 
collaboration \cite{Abbiendi:2001cs}. For centre of mass energies from 189 to 209 GeV the following limits were obtained: 
\begin{equation}
    \sigma(e^+ e^- \to \ell \ell') < 166 \text{ fb}\,,
\end{equation}
for $\ell \ell'=\mu \tau$. The limits on $e\mu$ are slightly more constraining, however such final states cannot be produced at leading order within our model.
The leading diagram involving $H_2,\,\tilde H_2$ which leads to LFV probed at LEP-II is depicted in fig.~\ref{fig:LFV} (c). For $m_{H_2,\tilde H_2}=140$ GeV, the contributing scattering cross section is of order yoctobarn (i.e.\ $10^{-9}$ femtobarn, also referred to as a {\it shed}), which is far below the experimental sensitivity.
\medskip

\begin{figure}
\begin{minipage}{0.49\textwidth}
\begin{center}\includegraphics[width=0.8\textwidth]{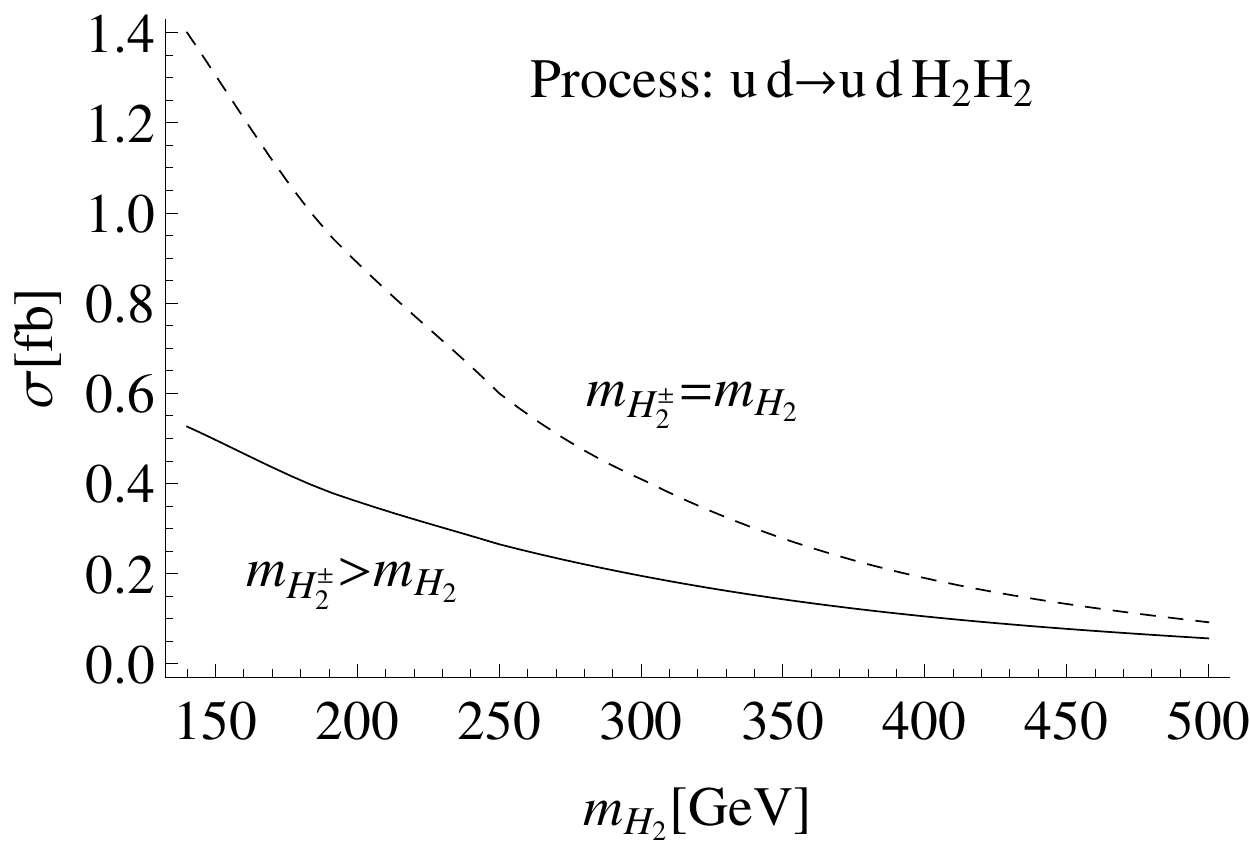}\end{center}
\end{minipage}
\begin{minipage}{0.49\textwidth}
\caption{Cross section of the dominating process contributing to the production of $H_2$ pairs at the LHC, for $\sqrt{s}=7$ TeV. The dashed black line denotes the particular model parameter setting of $m_{H_2^\pm} = m_{H_2}$, while the solid black line denotes the irreducible contribution from the gauge interactions of the $H_2$ field.}
\label{fig:h2production}
\end{minipage}
\end{figure}

It is worthwhile to consider LFV signatures of the ${\mathbb Z}_2$-even scalars at the LHC. The fields $H_2$ and $\tilde H_2$ can be produced via vector boson fusion\footnote{Note that the interactions of the scalar sector with the quarks has not been defined within our model.} and decay into lepton flavor violating dibosons:
\begin{equation}
p\, p \to j\,j\, H_2\,H_2\,,\qquad \text{with LFV decays} \qquad H_2 \to \ell\,\ell'\,,
\end{equation}
where $j$ is an energetic jet, and the same process exists for the replacement $H_2\to\tilde H_2$.
The dominant sub-process is given by $u\,d \to d\, u\, H_2\,H_2$ (and analogously for $\tilde H_2$). We show its cross section in fig.~\ref{fig:h2production} for $\sqrt{s}=7$ TeV. The numerical values were obtained with CalcHEP \cite{Pukhov:1999gg} and the CTEQ6 parton distribution functions \cite{Stump:2003yu}.
With the present integrated luminosity at the LHC of ${\cal L}$ = 20.3 fb$^{-1}$ a total number of LFV events of $\sim 10$ can be expected for each, the $H_2$ and the $\tilde H_2$. The combinations of lepton flavour thus produced are $e \tau$ and $\mu \tau$. Note that the expected number of events for the lepton-flavour combination $e \mu$ is suppressed by the ratio $m_\mu^2/m_\tau^2$.

Recently the direct search for lepton-flavour-violating Higgs boson decays of the CMS collaboration yielded an excess of 2.5$\sigma$ in the Higgs-to-$\mu$-$\tau$ channel \cite{CMS:2014hha, Khachatryan:2015kon}.
The LFV excess resides in the energy bin from 120 to 140 GeV and was fitted as an LFV decay of the SM Higgs boson. 
It is straightforward to consider this LFV excess being caused by the $H_2$ itself, which implies $m_{H_2,\tilde H_2}$ to be at the lower bound of 140 GeV, which implies in turn that MEG is about to observe the first $\mu \to e \gamma$ events. 
Note that in order to account for the observed $h \to \mu \tau$ excess, more than ${\cal O}(10)$ LFV events are necessary - the 10 events estimate presented above was considering only vector boson fusion processes. Interestingly, the 2-jet LFV events reported in \cite{Khachatryan:2015kon} are $4.1 \pm 1.1|_{\mu \tau_h} \pm 0.9|_{\mu \tau_e}$, which is around the order of the expected number of vector boson fusion events in our model.

\subsubsection{Effects of a Perturbation of the VEV Alignment}
\label{Ap:misalign}

The bounds above apply strictly speaking only if $\sigma_{3,(2)}=\sigma_{4,(2)}=0$ in $V_{\chi \phi}$ (see eq.~(\ref{ap:Vchiphi})). While this may be possible at the classical level, quantum corrections, e.g.\ the ones induced by the terms in Appendix~\ref{AppendixC}, can render these parameters non-zero, which deforms the VEV of $\Phi^{(2)}$, as shown in eq.~\eqref{eq:VEVs}, to
%
\begin{align}\label{eq:perturbedVEV}
    \left\langle \Phi^{(2)} \right\rangle &= \frac{v_\mathrm{EW}}{\sqrt{6}}
        \begin{pmatrix} 
          0 & 0 & 0 \\
          1 - \delta - \ci \sqrt{3} \, \delta^\prime & 1 + 2 \,\delta & 1 - \delta + \ci \sqrt{3} \,\delta^\prime 
        \end{pmatrix}\,.
\end{align}
In the following, we assume the parameters $\delta$ and $\delta^\prime$ to be small in order to allow for a perturbative treatment. They are given to leading order in $r$ and $\sigma_{3,(2)}$, $\sigma_{4,(2)}$ by
\begin{align}
    \delta &\approx \frac{
        \sigma_{3,(2)} + \sigma_{4,(2)}
      }{
        2 \sqrt{2} \, (\lambda_{4,(2)} + \lambda_{5,(2)})
      } \, \frac{1}{r} \,, \\
    \delta^\prime &\approx \frac{
        \sigma_{3,(2)} \, \lambda_{5,(2)} + \sigma_{4,(2)} \, (\lambda_{4,(2)} + 2 \,\lambda_{5,(2)})
      }{
        2 \sqrt{2} \, \sigma_{2,(2)} \, (\lambda _{4,(2)} + \lambda_{5,(2)})
      } \,,
\end{align}
where we assumed $\sigma_{3,(2)}$ and $\sigma_{4,(2)}$ to be real for simplicity.
The requirement of $\delta,\,\delta^\prime$ being small demands the parameters $ \sigma_{3,(2)}$, $\sigma_{4,(2)}$ to compensate a factor $1/r$. Furthermore, due to $r \ll 1$, the subleading perturbation proportional to $\delta^\prime$ can be neglected.

With the perturbed VEV alignment in eq.~\eqref{eq:perturbedVEV}, the charged lepton mass matrix of eq.~\eqref{eq:chargedleptonmassmatrix} is not diagonal after rotation with $V^l$ and requires additional rotations $\theta_{ij} \sim {\cal O}(\delta)$ in the left-handed sector\footnote{While this might change the conclusions made for the flavour model this paper is based on, we will not go into more detail on this particular issue and assume it non-problematic.} and $\theta_{ij} \sim {\cal O}(\delta) \, y_i/y_j$ in the right-handed sector. As a result the zeros in the Yukawa matrices $Y_1$, $Y_2$ and $\tilde{Y}_2$ (in the mass eigenbasis) are corrected to (up to leading order in $\delta$)
\begin{equation}\label{eq:y1perturbed}
Y_1 = \frac{1}{\sqrt{2}}\begin{pmatrix} 
        y_e & {\cal O}(\delta) \, y_\mu & {\cal O}(\delta) \, y_\tau \\
        {\cal O}(\delta) \, y_e & y_\mu & {\cal O}(\delta) \, y_\tau \\
        {\cal O}(\delta) \, y_e& {\cal O}(\delta) \, y_\mu & y_\tau
    \end{pmatrix}
    \times \left(1 - \left|\frac{\sigma_{1,(2)}}{4d_1+d_2}\right|^2 r\right)\,,
\end{equation}
and
\begin{equation}\label{eq:y2perturbed}
Y_2 = \frac{1}{2}
    \begin{pmatrix} 
        {\cal O}(\delta) \, y_e & y_\mu \omega^2 & y_\tau \omega \\
        y_e \omega & {\cal O}(\delta) \, y_\mu & y_\tau \omega^2 \\
        y_e \omega^2 & y_\mu \omega & {\cal O}(\delta) \, y_\tau
    \end{pmatrix}\,,
\end{equation}
with analogous structure for $\tilde{Y}_2$. 

The perturbed Yukawa matrices above allow for additional flavour-changing vertices for $H_1$ and flavour-conserving vertices for $H_2$ and $\tilde{H}_2$, such that non-zero contributions to the processes $\ell_i \to 3 \ell_j$ arise. These can be estimated as
\begin{equation}
  \frac{\text{BR}(\ell_i \to 3\ell_j)}{\text{BR}(\ell_i \to \ell_j \nu \bar{\nu})}
    \approx \frac{m_i^2 m_j^2}{m_H^4} \, \delta^2 \,,
\end{equation}
where $H$ is the scalar boson mediating the decay as described by diagrams of the form of fig.~\ref{fig:LFV} (a). Assuming $H = H_1$ with $m_{H_1} \approx 126$ GeV, the most stringent bound on $\delta$ comes from $\tau \to 3 \mu$ and is given by $\delta < 30$. 
Since $\delta<1$ in order to allow for the perturbative treatment, these kind of processes do not give meaningful bounds on the misalignment at present.

Likewise, when considering the branching ratios BR($\ell_i \to \ell_j \gamma$) as described by fig.~\ref{fig:LFV} (b) and similar, we find further contributing diagrams. Compared to the unperturbed VEV alignment, only the two processes $\tau \to \mu \gamma$ and $\tau \to e \gamma$ receive potentially significant contributions, as it is now possible to have tau leptons in the loop, with large Yukawa couplings to the scalars. 
Said contributions to the branching ratio are ``enhanced'' by a factor of $m_\tau^2/m_\mu^2\, {\cal O}(\delta^2) \sim 280 \, \delta^2$. Comparing the result for an unperturbed VEV from eq.~\eqref{eq:taugammadecays} with the experimental bounds of ${\cal O}(10^{-8})$, it shows that no bounds on $\delta$ can be obtained from this type of processes either.

Physical processes where the perturbation of the VEV alignment could lead to observable effects, are the LFV decays of the $H_1$ boson. 
In particular the branching ratio $\text{BR}(h \to \mu \tau)$ can be estimated as ${\cal O}(\delta^2)\text{BR}(h\to\tau\tau)$. This allows to use the observed excess of $h\to \mu \tau$ events at the LHC with a branching ratio of 0.84$^{+0.39}_{-0.37}\%$ to determine $\delta \sim {\cal O}(0.1)$. In this case, it is not necessary any more to have the masses of $H_2$ and $\tilde H_2$ close to the MEG bound.
For a more comprehensive analysis of the effect of a perturbed VEV alignment in a similar framework, also considering the $h\to \mu \tau$ excess, see ref. \cite{Heeck:2014qea}.

\subsubsection{Oblique Parameters \label{sec:oblique}}
The difference between the masses of the real, pseudo-real and charged fields from the previous section can be constrained via their contributions to the oblique parameters $S$, $T$, $U$. This has been studied for instance in refs.~\cite{Grimus:2008nb,Toorop:2010ex}. 
The strongest constraint comes from the $T$ parameter. A recent fit of the SM parameters to a global data set yields \cite{Baak:2014ora}
\begin{equation}\label{eq:Tparameter}
T =  0.09 \pm 0.13\,.
\end{equation}
In order to assess the contribution to the $T$ parameter we consider a generation of a neutral scalar field $S$, a pseudoscalar field $\tilde S$ and a charged scalar field $S^\pm$.
We define the masses of the three fields by
\begin{equation}
    m_{S^\pm}  =  \rho\, m_S 
    \qquad \text{and} \qquad 
    m_{\tilde S}  =  \tilde \rho\, m_S\,.
\end{equation}
We can approximate the contribution of one generation of scalar fields, $S,\,\tilde S,\, S^\pm$, to the $T$-parameter by
\begin{equation}\label{eq:TAi}
\delta T_{S} \approx  -\frac{1}{6\pi s_W^2}\, \left[(\rho-1)^2-(\tilde\rho-1)^2 - (\rho-\tilde\rho)^2\right] \,\frac{m_S^2}{m_W^2}\,,
\end{equation}
with $s_W$ being the sine of the weak mixing angle and $m_W$ the mass of the $W$ boson.
We observe that for $\tilde \rho = \rho$ or $\tilde \rho = 1$ the contributions of the three fields cancel exactly.
Furthermore, it is interesting to note that $\delta T_S > 0$ for $\rho > \tilde \rho$ and $\delta T_S < 0$ for $\rho < \tilde \rho$, which allows the possibility of cancellations between the contributions from different generations of scalars.

\subsubsection{Diphoton Decays of the Higgs Boson}
\begin{figure}
\begin{center}
\includegraphics[width=0.6\textwidth]{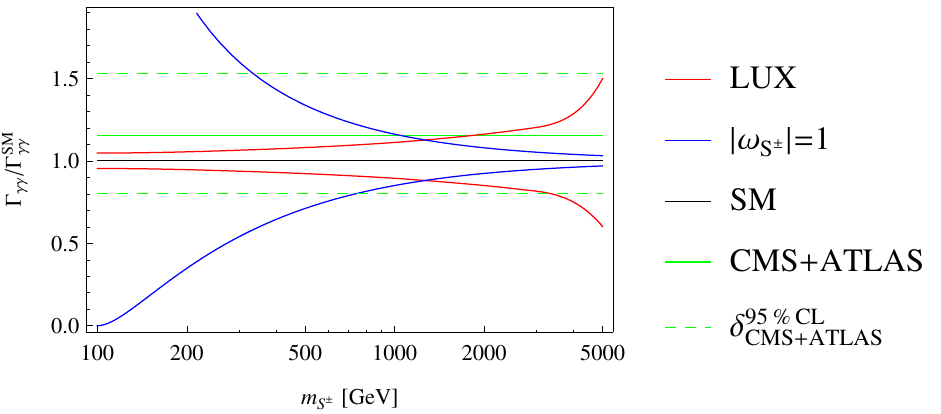}
\end{center}
\caption{Higgs to diphoton decay width, normalised over the SM prediction, in the presence of one extra charged scalar boson of mass $m_{S^\pm}$ and coupling to the Higgs boson $\omega_{S^\pm}$. 
The green line denotes the combined experimental precision from CMS \cite{Chatrchyan:2013iaa,Chatrchyan:2013mxa, Khachatryan:2014ira} and ATLAS \cite{Aad:2013wqa, Aad:2014eha}. 
The blue line denotes the maximum deviation for $|\omega_{S^\pm}|=1$, the red line indicates the limit set by the LUX collaboration on the spin-independent-DM-nucleon cross section, which may be related to the Higgs coupling of the charged ``sibling'' of the DM particle, see section~\ref{sec:directDM}.}
\label{fig:Hgg}
\end{figure}
The theory prediction for the process $H_1 \to \gamma\gamma$, where $H_1$ is the SM-like Higgs boson in our model, is given by:
\begin{equation}
    \Gamma_{\gamma \gamma} = \frac{G_F \alpha^2 m_{H_1}^3}{128 \sqrt{2} \pi^3} 
    \left| 
        {\cal A}_W (\tau_W) + 
        \sum_{\rm fermions} N_{c,f} Q_f^2 {\cal A}_F (\tau_f) + 
        \sum_{X} Q_{X}^2 {\cal A}_{X} (\tau_{X}) 
    \right|^2 \,, \\
\end{equation}
with $\tau_x =m_{H_1}^2/4 m_x^2$, $N_{c,x}$ being the number of colour states (3 for quarks, 1 for leptons), $Q_x$ is the electric charge of the particle in the loop, and the amplitudes $A (\tau)$ are functions that depend on the spin and couplings to the Higgs of the particle running in the loop.

The amplitude ${\cal A}_{X}$ denotes contributions from the electrically charged non-SM scalar fields, in our model given by $X=A_{i=1,2,3}^\pm,\,H_{j=2,3}^\pm$, with electric charge $|Q_{X}|= 1$.
We define the fields $A_{i=1,2,3}^\pm$ in section~\ref{sec:A} below, as the mass eigenstates of the ${\mathbb Z}_2$-odd fields $\eta^\pm_{i=1,2,3}$.
The amplitude ${\cal A}_f$ for the field $f=F,W,S$ being a fermion, vector boson or scalar, respectively, can be expressed as~\cite{Spira:1995rr}
\begin{eqnarray}
{\cal A}_F (\tau) &=& \frac{2}{\tau^2} \left( \tau + (\tau-1) f (\tau) \right)\,, \\
{\cal A}_W (\tau) &=& - \frac{1}{\tau^2} \left( 2 \tau^2 + 3 \tau + 3 (2 \tau - 1) f (\tau) \right)\,, \\
{\cal A}_S (\tau) &=& - \frac{1}{\tau^2} \left( \tau - f (\tau) \right)\,;
\end{eqnarray}
with the function
\begin{equation}
    f (\tau) = \begin{cases}
        \mbox{arcsin}^2 \sqrt{\tau}  & \tau \leq 1 \\
        - \frac{1}{4} \left[ \log \frac{1+\sqrt{1-\tau^{-1}}}{1-\sqrt{1-\tau^{-1}}} - i \pi \right]^2 & \tau > 1 
    \end{cases} \,.
\end{equation}
Note that this definition of the ${\cal A}_f$ includes the SM relation between the mass of a particle and its coupling to the 125.7 GeV Higgs boson $H_1$.

We start by considering a generic electrically charged scalar boson $S^\pm$, with mass $m_{S^\pm}$ and an interaction term with the SM-like Higgs boson of the form: $\omega_{S^\pm}\,v_\mathrm{EW}\, H_1\,S^\pm S^\mp$.
The contribution of the scalar field $S^\pm$ to the SM-Higgs-to-diphoton decay width can be expressed as
\begin{equation}
    {\cal A}_{S^\pm}(\tau_{S^\pm}) = \omega_{S^\pm} \, {246 \text{ GeV} \over m_{S^\pm}} \, {\cal A}_S(\tau_{S^\pm})\,,
\label{eq:scalaramplitude}
\end{equation}
where the product on the right-hand side of eq.~\eqref{eq:scalaramplitude} takes into account the deviation from the SM relation between coupling and mass of a particle. 

We show the ratio of the decay rate $\Gamma_{\gamma\gamma}$ -- for the case where the SM is extended with one extra charged scalar field $S^\pm$ -- over the SM prediction $\Gamma_{\gamma\gamma}^\mathrm{SM}$ in fig.~\ref{fig:Hgg}.
Notice that an enhancement requires positive interference with the top quark, which occurs for $\omega_{S^\pm}<0$.
It shows that even for the coupling between the scalar $S^\pm$ and the Higgs boson being order unity, the deviations from the SM prediction are compatible with the experimental result for $m_{S^\pm}> 200$ GeV. For $m_{S^\pm}$ on the TeV scale and with couplings order one the induced deviations remain on the percent level. This is beyond the reach of the LHC even with the luminosity upgrade.

As we have mentioned above, in our model the electrically charged scalar fields relevant for the diphoton decay width of the Higgs boson are the fields $H_{j=2,3}^\pm$ and $A^\pm_{i=1,2,3}$ (see section~\ref{sec:A} for the definition). 
In fig.~\ref{fig:Hgg} we include the LUX constraint on the modulus of the coupling between the (electrically neutral) $A_{\DM}$ field and the Higgs boson $|\omega_{\DM}|$, which might be related to $\omega_i^\pm$, the coupling between $H_1$ and $A_i^\pm$. 

The deviations from the SM prediction due to charged scalar fields with masses around the TeV scale could be studied for instance at future lepton colliders. The following precision can be achieved for one interaction point after one year of data taking: 35\% at the International Linear Collider \cite{Baak:2013fwa}, 8\% at the Circular Electron-Positron Collider \cite{Ruan:2014xxa} and 3\% at the Future Circular Collider in the electron-positron mode \cite{Gomez-Ceballos:2013zzn}.

\section{${\mathbb Z}_2$-odd Scalar Sector \label{sec:A}}

\subsection{Physical Field Content}

Analogously to the previous section, we define the physical DM fields $A_i,\,\tilde A_i$ and $A^\pm_i,\,i=1,...,3$ through the eigenstates of the corresponding mass matrices of the ${\mathbb Z}_2$-odd fields $\eta,\tilde \eta,\eta^\pm$ as defined in eq.~\eqref{eq:fielddefinition}. Note that the mass eigenstates are admixtures of both $\eta$ and $\tilde \eta$.

We verified that no new gauge vertex is generated through the unitary transformation into the physical basis. Furthermore, since the gauge interactions of the $\eta$ and $\tilde \eta$ are identical up to the complex phase i, the total annihilation cross section for two ${\mathbb Z}_2$-odd into two ${\mathbb Z}_2$-even fields remains unchanged.
For $r = 0$ and with the VEV alignment from eq.~\eqref{eq:VEVs}, the mass eigenvalues for all $SU(2)_L$ components of the three ${\mathbb Z}_2$-odd scalar fields $\Phi^{(1)}_i$ are given by:
\begin{equation}\label{eq:A0jmasses}
\begin{split}
m_{A_1}^2 & = 2 \mu _1^2 + (-\left.\text{Re[}\sigma _{3,(1)}\right]+2 \sigma _{1,(1)})\, v_\chi^2 \,,  \\
m_{A_2}^2 & = 2 \mu _1^2 + \frac{1}{2} \left(\left.\text{Re[}\sigma _{3,(1)}\right]+4 \sigma _{1,(1)}+2 \sigma _{2,(1)}+2 \sigma _{4,(1)}\right)\, v_\chi^2 - \frac{1}{2}\delta m^2_A\,,  \\
m_{A_3}^2 & = 2 \mu _1^2 + \frac{1}{2} \left(\left.\text{Re[}\sigma _{3,(1)}\right]+4 \sigma _{1,(1)}+2 \sigma _{2,(1)}+2 \sigma _{4,(1)}\right)\, v_\chi^2 + \frac{1}{2}\delta m^2_A\,,  \\
m_{A_i^\pm}^2 & = m_{\tilde A_i}^2 = m_{A_i}^2
\end{split}
\end{equation}
with $\delta m^2_A = v_\chi^2 
\sqrt{3 \left.\text{Im[}\sigma _{3,(1)}\right]{}^2+4 \left(\sigma _{2,(1)}^2-\sigma _{2,(1)} \sigma _{4,(1)}+\sigma _{4,(1)}^2\right)}$.
In the limit of $r=0$, the eigenvalues from the mass matrices for the ${\mathbb Z}_2$- and electrically charged scalar fields and the two neutral fields are identical and they form complete $SU(2)_L$ doublets, i.e.\ all three components are a generation of mass-degenerate fields.
The couplings to the $\Phi^{(2)}$ field break this degeneracy, however. The expressions are too cumbersome to be displayed here in a useful way. Fortunately for the analysis to follow, it suffices to state that the masses of the fields $A_i,\,\tilde A_i,\,A_i^\pm$ with $i=1,2,3$ are all different, and that the relevant mass differences are of the order of $v_\mathrm{EW}$.

\subsection{Effective Parametrisation \label{sec:parametrisation}}
As can be seen in eq.~\eqref{eq:A0jmasses}, the masses of the three $A_i$ fields can be chosen independently.
The same is true for the mass splittings between each generation $A_i$ and its ``siblings'' $\tilde A_i,\,A_i^\pm$, which is induced by the (small) VEV of $\Phi^{(2)}$.\footnote{As mentioned already, the resulting expressions are very cumbersome and not very enlightening, which is why we do not to present them here.}
We will thus use the three physical masses $m_{A_i}$, and the six mass-splittings between the $A_i$ and the $\tilde A_i,\,A_i^\pm$, respectively, as free parameters. We define the latter for $i=1,2,3$ by:
\begin{equation}\label{eq:mass-splitting}
  \rho_i = \frac{m_{A_i^\pm} - m_{A_i}^{}}{m_{A_i}^{}}\,, 
\qquad
  \tilde \rho_i = \frac{m_{\tilde A_i} - m_{A_i}^{}}{m_{A_i}^{}}\,.
\end{equation}
Other important parameters are the effective couplings between the $A_i,\,\tilde A_i,\,A_i^\pm$ and the SM-like Higgs field $H_1$. We define this effective coupling for $A_i$:
\begin{equation}
\omega_i = \frac{\partial^4 V }{(\partial A_i)^2 (\partial H_1)^2} \,,
\end{equation}
with $V$ being the scalar potential from eq.~\eqref{eq:potential}. Analogously we can define the couplings $\tilde \omega_i,\,\omega_i^\pm$.
Especially $\tilde \omega_i$ and $\omega_i^\pm$ are related to the mass splitting parameters $\rho_i$ and $\tilde \rho_i$, respectively, because they stem from the same term in the potential $V_{\phi\phi}$, but they have no effect on the phenomenological study in the following.

We note the gauge interactions are identical for the $A_{i=1,2,3}$ and $\tilde A_{i=1,2,3}$ up to a complex phase and the couplings between the $A_{i=1,2,3}$ or the $\tilde A_{i=1,2,3}$ and the $H_1$ depend on the parameters $\kappa_i$ in $V_{\phi\phi}$, which are independent from the parameters that generate the masses $m_{A_i,\tilde A_i}$.
Therefore the choice of a specific DM candidate from the three generations is without loss of generality.

\subsection{Phenomenology \label{pheno}}
In this section we will explore the dark matter phenomenology of the above presented model with the effective parametrisation from section~\ref{sec:parametrisation}. 
In the limit of $r=0$ there are two mass-degenerate scalar DM candidates in generation $i$, given by the electrically neutral fields $A_{i}$ and $\tilde A_{i}$, both with mass $m_{A_i}^{}$ and for $m_{A_i}^{}<m_{A_{j\neq i}}$. Depending on the contribution from $v_\mathrm{EW}$, two scenarios are given, the first one being defined by $m_{A_i}^{} < m_{\tilde A_i}$ and $m_{A_i}^{}<m_{A_{j\neq i}}^{}$ (scenario 1), and the second one being defined by $m_{A_i}^{} > m_{\tilde A_i}$ and $m_{\tilde A_i}^{}<m_{A_{j\neq i}}^{}$ (scenario 2).
The fields, the $A_{i=1,2,3}$ and the $\tilde A_{i=1,2,3}$ are linear combinations of the real and imaginary parts of $\Phi^{(1)}_{i=1,2,3}$, with the gauge couplings being identical up to a factor i that is associated to the $\tilde \eta_{i=1,2,3}$. The gauge couplings of the $A_{i=1,2,3}$ and $\tilde A_{i=1,2,3}$ are thus identical up to a complex phase. In the following we study the limiting  case, where $A_{i=1,2,3}$ consists exclusively in $\eta_{i=1,2,3}$ fields, while $\tilde A_{i=1,2,3}$ consists exclusively in $\tilde \eta_{i=1,2,3}$ fields. By a suitable mixing of real and pseudo-real fields a continuous interpolation between the two scenarios should be possible.

We identify the DM field with the electrically neutral ${\mathbb Z}_2$-odd field with the smallest mass and denote it by $A_{\DM}$. 
The other neutral field will be denoted by $\tilde A_{\DM}$: in scenario 1, $\tilde A_{\DM} = \tilde A_i$ while in scenario 2, $\tilde A_{\DM} = A_i$.
Furthermore we introduce the parameters $m_{\DM},\rho_{\DM},\tilde \rho_{\DM}$, which are given in the two scenarios by
\begin{equation}\label{eq:defscenario}
\begin{array}{clll}
\text{scenario 1:} & m_{\DM} = m_{A_i}\,, & \rho_{\DM} = \dfrac{m_{A^\pm_i}-m_{A_i}^{}}{m_{ A_i}}\,,& \tilde \rho_{\DM} = \dfrac{m_{\tilde A_i}-m_{ A_i}^{}}{m_{ A_i}}\,, \\[3ex]
\text{scenario 2:} & m_{\DM} = m_{\tilde A_i}\,, & \rho_{\DM} = \dfrac{m_{A^\pm_i}-m_{\tilde A_i}}{m_{\tilde A_i}}\,,& \tilde \rho_{\DM} = \dfrac{m_{A_i}^{} - m_{\tilde A_i}}{m_{\tilde A_i}}\,.
\end{array}
\end{equation}
Note that in scenario 1, $\rho_{\DM}$ and $\tilde \rho_{\DM}$ coincide with the definitions for $\rho_i$ and $\tilde \rho_i$ in eq.~\eqref{eq:mass-splitting}.
We note here, that even though $m_{\DM}$ is proportional to $v_\chi$, a cancellation of terms allows the mass-scale to be below the TeV scale despite the assumption in eq.~(\ref{eq:approximation}).

Despite $m_{H_2,\tilde H_2} \propto v_\mathrm{EW}$, we assume in the following that all the scalar non-DM fields have masses above $2\,m_{\DM}$. 
This is done in order to avoid a distortion of the total DM annihilation cross section by additional scalar resonances. In the present model and due to the total decay width being suppressed by the small lepton-Yukawa couplings, for instance the $H_2$ resonance would manifest as a very narrow peak in fig.~\ref{fig:scatterscan}, leading to smaller minimal values for $\Omega_\DM$ at $m_\DM = m_{H_2}/2$. However, since the quark Yukawa interactions are not defined in our model, neither the exact shape nor the magnitude of the scalar resonance is properly defined.

The right-handed Majorana neutrino $N_R$ is also a viable candidate for DM. 
The combination of the Yukawa coupling $y_\nu$ and the Majorana mass $M_N$ are required to match the light neutrino masses. In ref.~\cite{Hernandez:2013dta} it was found that for $y_\nu$ about the Yukawa coupling of the electron, $M_N = {\cal O}($TeV), while for larger values of $y_\nu$, $M_N$ needs to become larger as well.  
Therefore we shall always keep $M_{N} > m_{\DM}$ in the following.

\subsubsection{Constraints from Terrestrial Experiments}
The experimental bounds on the masses of charged and neutral scalar fields from LEP-I, which entail essentially the width of the $Z$ boson, are given by \cite{Agashe:2014kda}:
\begin{equation}\label{eq:massbound}
    m_{A_i,\,\tilde A_i,\, A_i^\pm} > m_Z/2 \text{ (LEP-I)}\,.
\end{equation}
Those bounds are stringent on all models that have particles with couplings to the $Z$ boson. Furthermore, lower bounds on the masses of charged and neutral scalar particles, denoted by $S^\pm$ and $S$, respectively, which decay into SM leptons have been derived from LEP-II data \cite{Agashe:2014kda}:
\begin{equation}\label{eq:massbound2}
    m_{S,\,S^\pm} > 80 \text{ GeV (LEP-II)}\,,
\end{equation}
The unbroken ${\mathbb Z}_2$ symmetry that stabilises the DM candidate also forbids the decays of the fields $A_i$, $\tilde A_i$ and $A_i^\pm$ into SM leptons, such that the bound in eq.~\eqref{eq:massbound2} does not apply.

The OPAL collaboration has published exclusion bounds at 95\% confidence level for supersymmetric chargino and neutralino particles at LEP-II, see e.g.\ refs.~\cite{Abbiendi:2003ji,Abbiendi:2003sc}, at centre of mass energies from 189 to 209 GeV. The bound can be expressed as
\begin{equation}\label{eq:Opalbound}
    \sigma(e^+ e^- \to A_i^\pm A_i^\mp) < 0.1 \text{ pb} \qquad \text{for } \qquad \sqrt{s} = 189 \text{ GeV}\,.
\end{equation}
Using CalcHEP \cite{Pukhov:1999gg} we get for $m_{A_1^\pm} = m_W$  a cross section for the above process of 0.16 pb. The bound in eq.~\eqref{eq:Opalbound} can therefore be interpreted as a lower bound on the charged DM mass $m_{A_{i}^\pm} > 120$ GeV for $m_{A_i}= m_\DM$.
However, it is obvious from the exclusion plots in \cite{Abbiendi:2003ji} that for small mass splitting, i.e.\ $\rho_{\DM}\simeq 1$ the above bound does not apply.
In this corner of the parameter space the decay products of the charged $A_{\DM}^\pm$ become very soft and can escape detection due to the selection cuts from the experimental analyses. Furthermore, assuming that a number of observable events are being produced, the cross section is too small to produce a significant excess over the SM backgrounds.
Subsequently we will allow for $\rho_\DM$ and $\tilde \rho_\DM$ up to 1.2.
This leads us to the two distinct possibilities for DM with masses below $m_W$: either the mass-splitting between the neutral field $A_\DM$ and the charged field $A_\DM^\pm$ needs to be large enough such that $m_{A_\DM^\pm} > m_W$, or $\rho_\DM,\,\tilde \rho_\DM \leq 1.2$.
\medskip

As discussed in section~\ref{sec:oblique}, the $T$ parameter is very sensitive to additional scalar doublets. Among the ``dark sector'' fields the mass-splitting parameters $\tilde \rho_i,\,\rho_i$ can be set such that the summed contributions of the second and third generation remains compatible with the experimental bounds.
The mass-splitting parameters $\rho_\DM$ and $\tilde \rho_\DM$ are very important to control the produced relic density, as we will show below. We will use eq.~\eqref{eq:TAi} therefore to constrain the parameter space automatically, unless stated otherwise.

The constraints from experimental tests on lepton flavour violation also apply to the fields from the ``dark sector''. In particular the ${\mathbb Z}_2$-odd scalar fields contribute to the process $\mu \to e \gamma$ and are thus subject to the strong constraint from the MEG collaboration, cf. fig.~\ref{fig:LFV} (b). Compared to the Higgs-like sector those constraints affect the combination of the masses $m_{A_i,\tilde A_i,A_i^\pm}$, the Yukawa coupling $y_\nu$ and the Majorana mass $M_N$. In practice given that we are assuming $y_\nu \sim y_e$ and $M_N \sim$ TeV \cite{Hernandez:2013dta}, and since these quantities are only constrained through the observed light neutrino masses (which gives a bound on the combination), it is not possible to derive meaningful bounds on the masses $m_{A_i,\tilde A_i,A_i^\pm}$ through the bound obtained from the MEG constraint alone.

\subsubsection{Dark Matter Abundance}
\begin{figure}
  \begin{center}
  \includegraphics[width=0.7\textwidth]{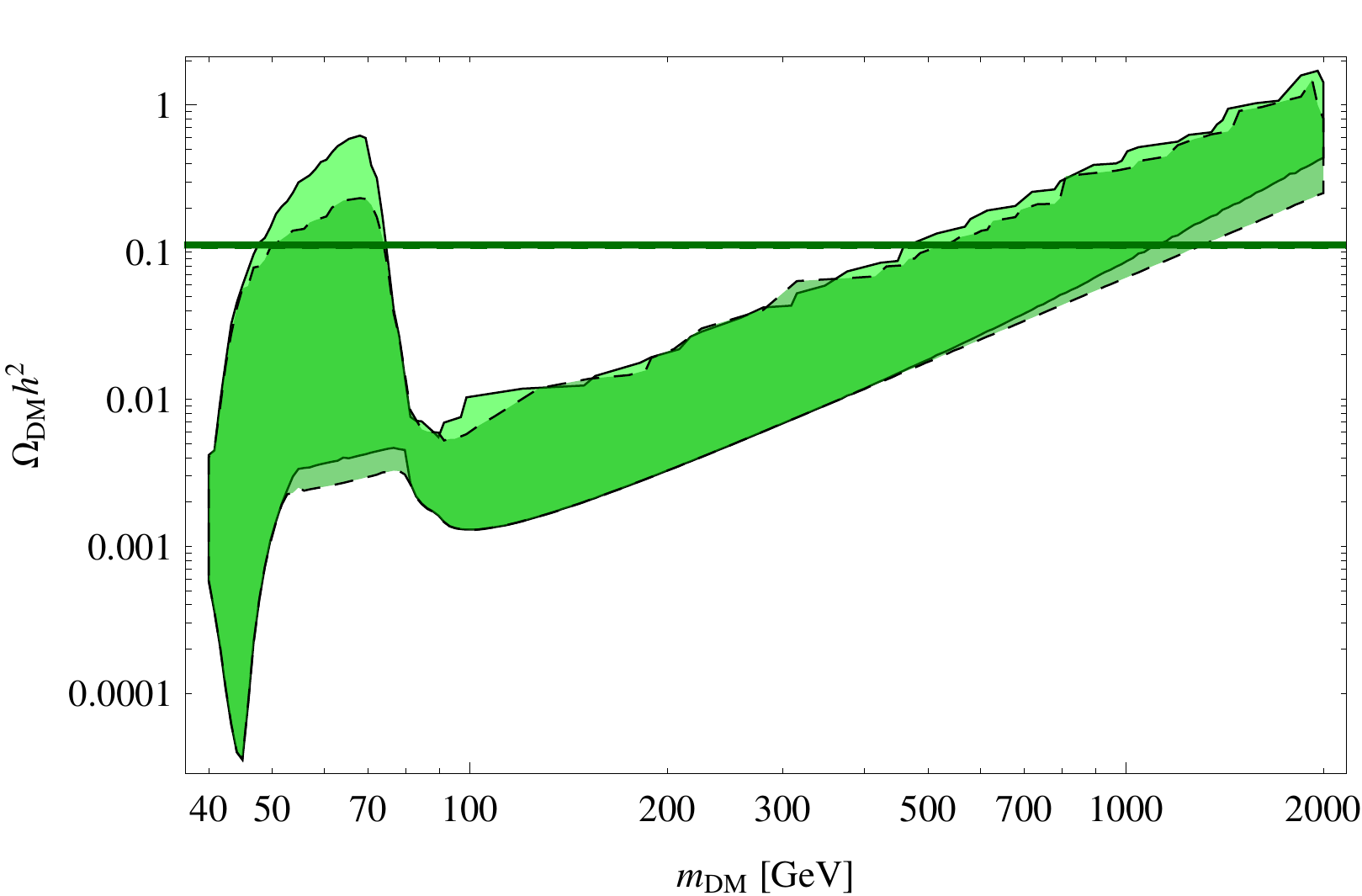}
  \end{center}
  \caption{Numerical results for the DM relic density $\Omega_{\DM}h^2$ over the DM mass $m_{\DM}$ from a scatter scan over the mass and mass-splitting parameters of the ${\mathbb Z}_2$-odd fields. The considered parameter space is defined in eq.~\eqref{eq:scatterscan}. The dark green line denotes the observed DM relic density \cite{Amsler:2008zzb}. 
The light and the dark green (non-dashed and dashed) areas denote the scenario 1 and 2, where the DM is given by the lightest $A_i$ and $\tilde A_i$, respectively. The constraints from the $T$ parameter on the mass-splitting are applied in both cases. 
In the figure, $m_{H_{i\geq2}}\geq 2\,m_{\DM}$ is assumed.}
  \label{fig:scatterscan}
\end{figure}

The relic density in this section as well as the DM-nucleon-interaction cross sections in the next section are computed with the program MicrOMEGAs \cite{Belanger:2004yn,Belanger:2014vza}. For all the numerical studies in this section we use the Tool for Parallel Processing in Parameter Scans (T3PS) \cite{Maurer:2015gva}.

In order to assess the limits for the two scenarios in our model, we compute the relic density $\Omega_{\DM}$ for the scenarios 1 and 2 and
for the following parameter ranges:
\begin{equation}\label{eq:scatterscan}
  \begin{array}{l}
    40 \text{ GeV} \leq m_{\DM} \leq 2 \text{ TeV}\,, \\
   1 < \rho_\DM,\, \tilde \rho_\DM \leq 1.2\,, \\
    m_{\DM} < m_{A_{i}\neq A_\DM},\,m_{\tilde A_{i}\neq \tilde A_\DM},\,m_{ A_{i}^\pm\neq A_\DM^\pm} \leq 10\, m_{\DM}\,, \\
    2 m_{\DM} < m_{H_{i\geq 2}} \,, \\
    \omega_\DM = 0\,, \\
    M_N = 10 \, m_\DM\,.
  \end{array}
\end{equation}
The resulting values for $\Omega_{\DM} h^2$ are shown in fig.~\ref{fig:scatterscan} by the light green/solid area for scenario 1, respectively by the dark green/dashed area for scenario 2.
The green horizontal line in the figure denotes the presently observed DM relic density \cite{Amsler:2008zzb}. 
The figure shows that in both scenarios parameter values exist, such that the relic density $\Omega_{\DM}$ may match the observation for the mass ranges below 100 GeV, and between 500 GeV $\leq m_{\DM} \leq 1.5$ TeV.

\begin{figure}
  \begin{center}
    \includegraphics[height=0.34\textwidth]{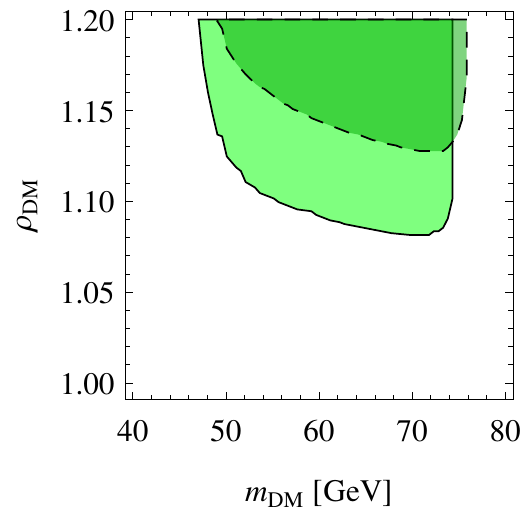}
    \hspace{-1.65cm}
    \includegraphics[height=0.34\textwidth]{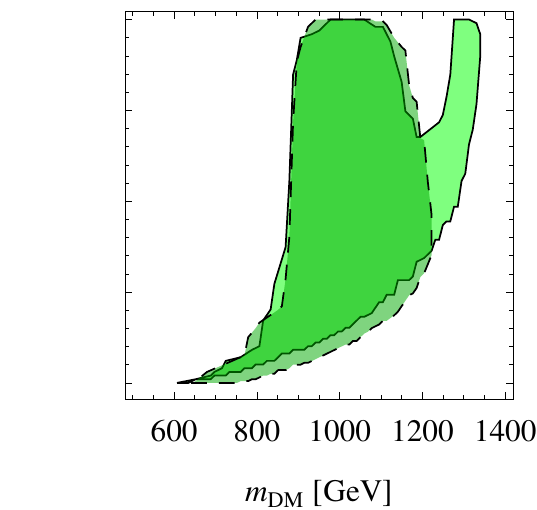}
    \includegraphics[height=0.34\textwidth]{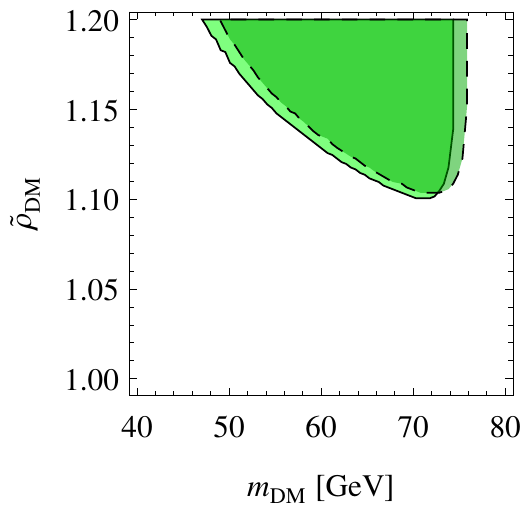}
  \end{center}
  \caption{{\it Left:} The relative mass-splitting parameter $\rho_{\DM}$ (as defined in eq.~\eqref{eq:defscenario}) over the DM mass. The light and dark green (non-dashed and dashed) areas denote resulting values for the relic density $\Omega_{\DM}$ in agreement with observation for scenario 1 and 2, respectively. {\it Right:} Mass splitting parameter $\tilde \rho_\DM$ between the two neutral DM fields $A_i$ and $\tilde A_i$ for $m_{A_i}< m_{A_{j\neq i}}$, over the DM mass. The colour-code is the same in both panels. The constraints from the $T$ parameter on the mass-splitting are applied in both cases.
In this figure, $m_{H_{i\geq2}},\,m_{A_{i\geq2}} =  3$ TeV has been used.}
  \label{fig:omega}
\end{figure}

We consider the mass ranges for $A_\DM$ and $\tilde A_\DM$, where the abundance constraint can be met,
and conduct two new parameter scans for each scenario with a higher resolution around the areas of interest. 
We set the masses of the $i=2,3$ DM generations to 3 TeV and the DM-Higgs coupling $\omega_{\DM}$ to zero, which is identical to $\omega_1$ and $\tilde \omega_1$ in scenario 1 and 2, respectively.
In fig.~\ref{fig:omega} by the green areas, we show the ($m_{\DM},\,\rho_{\DM}$) and ($m_{\DM},\,\tilde \rho_{\DM}$) parameter space that is compatible with the observation of the relic density.
The dark/dashed and light green/solid areas denote scenarios 1 and 2, respectively.

For the mass $m_{\DM} \sim$ 1.4 TeV, the electroweak processes of the form $A_\DM\, A_\DM \to W^+ W^-,\, ZZ$ provide exactly the right cross section to satisfy the abundance constraint.
This demands the co-annihilations to vanish, which in turn implies mass splitting $\rho_{\DM} > 1.2$.
In the figure this results in a confined mass range for the DM particle, which is comparable in both scenarios.
However, since $\Omega_{\DM}$ grows proportional to $m_{\DM}$, the abundance constraint can be met by increasing $\omega_{\DM}$ for $m_\DM > 1.4$ TeV.
The DM candidates from the two scenarios considered here have a mass that is confined to one of the two intervals:
\begin{equation}
    m_{A_1} \in \begin{cases} \text{47 GeV to 74 GeV}  \\ \text{600 GeV to 3.6 TeV} \end{cases}, 
        \qquad 
    m_{\tilde A_1} \in \begin{cases} \text{50 GeV to 74 GeV}  \\ \text{620 GeV to 3.6 TeV}\,, \end{cases}
\end{equation}
where the upper bound on the mass comes from the perturbative limit $\omega_i,\,\tilde\omega_i \leq 1$ for all $i$, and requires the condition $\rho_{\DM}>1.2$ and $\tilde \rho_{\DM}>1.2$, i.e.\ suppressed contributions from the other ``dark sector'' fields.

\subsubsection{Experimental Limits from Direct Dark Matter Searches \label{sec:directDM}}
\begin{figure}

	 \begin{subfigure}[b]{0.49\textwidth}
	\centering
                \includegraphics[width=0.9\textwidth]{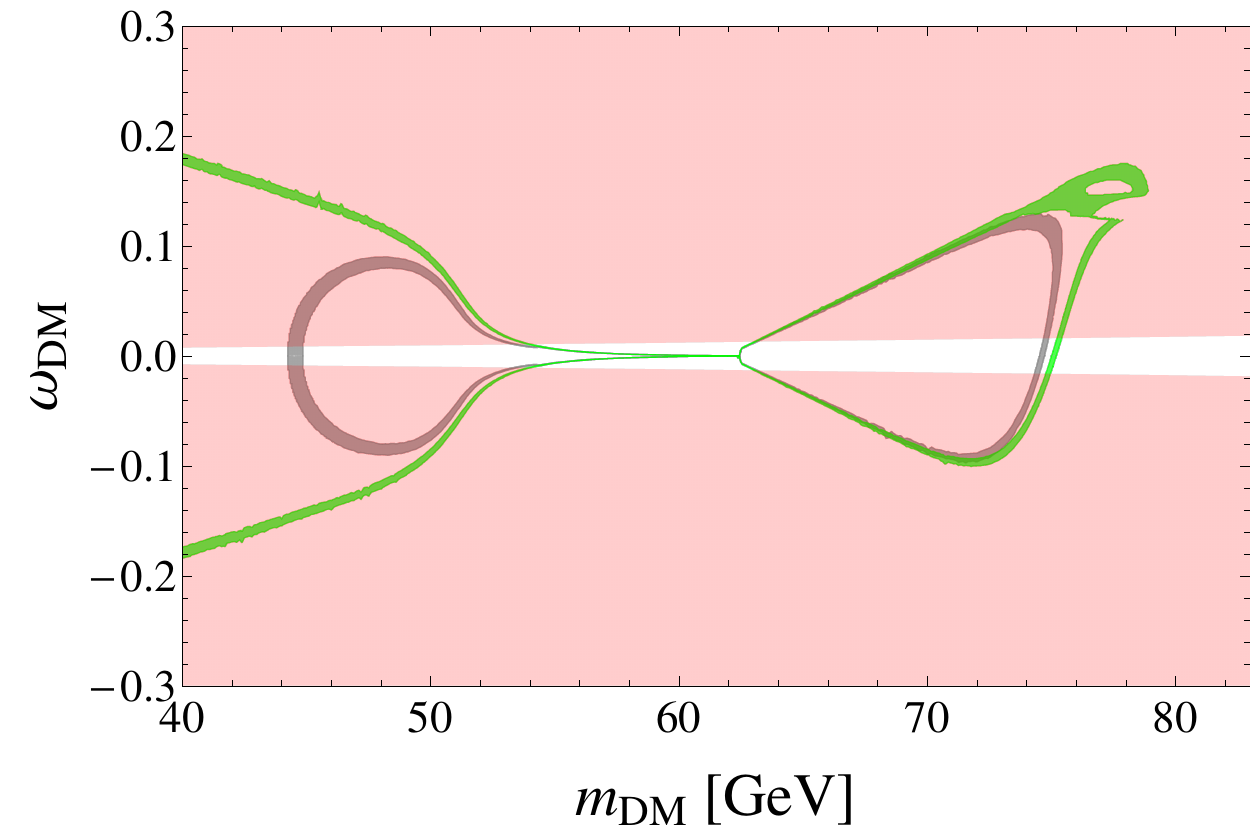}
                \caption{scenario 1}
        \end{subfigure}
        \begin{subfigure}[b]{0.49\textwidth}
	\centering
		\hspace{-1.cm}
                \includegraphics[width=0.9\textwidth]{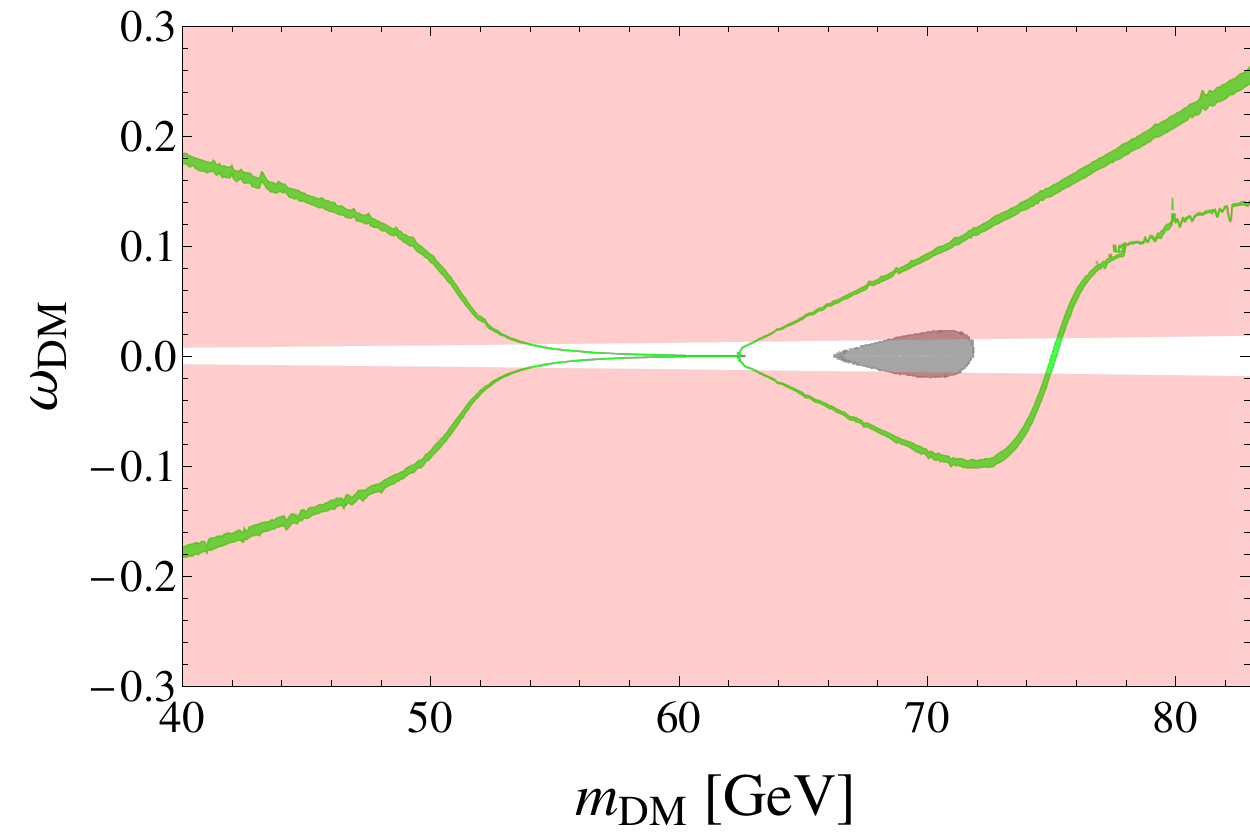}
                \caption{scenario 2}
        \end{subfigure}
\caption{
The {\it left} panel denotes scenario 1, where $A_{\DM} = A_i$, $m_{\DM} = m_{A_i}$ and $\omega_{\DM} = \omega_i$, while the {\it right} panel denotes scenario 2, where $A_{\DM} = \tilde A_i$, $m_{\DM} = m_{\tilde  A_i}$ and $\omega_{\DM} = \tilde \omega_i$, for $m_{A_i}< m_{A_{j\neq i}}$.
In both panels the green and grey bands denote the mass-splitting between $A_{\DM}$ and $A_{\DM}^\pm$, $\rho_{\DM}=2.0$ and $\rho_{\DM}=1.15$, respectively.
Parameter values inside the green and grey bands result in a relic density $\Omega_{\DM}$ that matches the observation. 
The red area denotes the exclusion constraints from the direct DM search experiment LUX \cite{Akerib:2013tjd}. 
In this figure, $\tilde \rho=2.0$ and all the other masses are chosen to be above 100 GeV.
}
\label{fig:higgscoupling}
\end{figure}
For $\rho_{\DM}, \tilde\rho_{\DM} > 1.2$, the observed value for the relic density can be used to constrain the $A_{\DM}$-Higgs coupling, parametrised by $\omega_{\DM}$.
We analyse both scenarios for the two values of $\rho_{\DM}=2.0$ and $\rho_{\DM} = 1.15$ and we set the mass splitting between the two neutral components $\tilde \rho_{\DM} = 2.0$. All the other masses are set to $10\times m_{\DM}$, to avoid co-annihilations.
We remark that regarding the numerical computation of the relic density both values of $\rho_\DM = 2$ and $\rho_\DM = 10$ lead to the same result. This is due to the Boltzmann suppression of the ``dark sector'' fields, which effectively removes their contribution to the total annihilation cross section for mass-splittings much larger than 20\%. The particular setting of the masses is thus somewhat arbitrary since it has no further impact on the result.
Note that the ``corner'' of the parameter space considered here does not lead to tension with the bounds on the $T$ parameter in eq.~\eqref{eq:Tparameter}.

The resulting constraints for $\omega_{\DM}$ are shown in the two panels in fig.~\ref{fig:higgscoupling}, where the green and grey band denote $\rho_{\DM} = 2.0$ and $\rho_{\DM}=1.15$, respectively. We observe that the allowed region for $\omega_{\DM}$ is being ``squeezed'' to smaller values, for $\rho_{\DM}$ approaching the critical value, which is given by the lower bound in fig.~\ref{fig:omega}. 
Also displayed in the figure are the LUX constraints, which are denoted by the red area, which affect the spin-independent $A_\DM$-nucleon cross section, $\sigma^{SI}_{\DM}$. The experimental upper bound is of order zb (10$^6$ zb $=$ 1 fb or zb = 1000  {\it sheds}) for $m_{\DM}$ between 10 and 100 GeV, and order 10 zb \cite{Akerib:2013tjd} for $m_{\DM}\sim 1$ TeV.
It is interesting that fig.~\ref{fig:higgscoupling} (b) shows the possibility to generate the correct DM relic density for scenario 2 also for $m_{\DM}>m_W$, which is due to a destructive interference in the total annihilation cross section that is not present in scenario 1.
Note, that the LUX bound relaxes for larger masses, such that for $m_{A_i}$ beyond 3 TeV a coupling $|\omega_{\DM}| \sim 1$ is possible.

The fact that the relic abundance can be generated in our model via the co-annihilations is visible in the ``squeezing'' of the contour for $\omega_{\DM}$, which occurs when $\rho_{\DM}$ becomes smaller than 1.2. 
Therefore the direct search constraints yield strong bounds on the DM interaction with the Higgs boson, but do not exclude the possibility of DM with 47 GeV $\leq m_{\DM} \leq$ 74 GeV.
\medskip

The spin-dependent interaction between the DM and the nucleon, with the corresponding cross section $\sigma^{SD}_{\DM}$, is mediated by the weak gauge bosons. 
Experimentally, $\sigma^{SD}_{\DM}$ is constrained especially by IceCube \cite{IceCube:2011aj} with upper limits of 0.1--1.0 fb, followed by the upper bound of 4.5 fb from Super-Kamiokande \cite{Tanaka:2011uf}, both valid for the mass range from $\sim 100$ GeV to 1 TeV. 
For masses below 100 GeV, the best upper bounds to the spin-dependent DM-nucleon interaction is given with $\sim 4$ fb from COUPP \cite{Behnke:2012ys} and SIMPLE \cite{Felizardo:2011uw}. 
The interaction between the field $A_\DM$ and the nucleon takes place only at the loop level via exchange of at least two $W$ or $Z$ bosons, and the cross section is suppressed by $\alpha^2_\mathrm{EW}$ compared to the usual tree-level expression involving the exchange of a single vector boson \cite{Fischer:2011zz}, which yields a typical cross section ${\cal O}(0.1)$ fb, for weak cross sections ${\cal O}$(pb).

\subsubsection{Signatures for Dark Matter at Hadron Colliders}
Direct searches for signatures of dark matter with charged components at the LHC have been considered e.g.\ in refs.~\cite{Barbieri:2006dq,FileviezPerez:2008bj}, and very recently also including the hadron-hadron mode of the Future Circular Collider (FCC-hh) planned at CERN \cite{Primulando:2015lfa}. 

The search for a signal in LHC data is very promising, provided the DM mass is within reach. In particular, event topologies with an imbalance in transverse momentum and one (or two) hadronic jet(s) or photon(s), respectively, may show an excess over the SM background.
In fact, for an appropriate choice of masses it might be possible to explain observed excesses in events with opposite sign di-lepton, jets and missing transverse momentum by ATLAS \cite{Aad:2015wqa} and CMS \cite{Khachatryan:2015lwa} - due to cascade decays of the three generations of $\mathbb{Z}_2$-odd scalars.
Furthermore, due to the specific vacuum alignment of our model, the interactions between fields of the second generation with those of the first and/or third generation are suppressed by the small parameter $r$, such that the second generation fields can have a long life time.
This in turn allows for signals with displaced vertices, which benefit from a very low SM background, which was recently considered in ref.~\cite{Primulando:2015lfa} in the context of an effective field theory approach.

Recently, the CMS collaboration has also published bounds on the spin-dependent-DM-nucleon interaction cross section from an analysis of 19.7 fb$^{-1}$ LHC data at $\sqrt{s}=8$ TeV \cite{Khachatryan:2014rra}. Therein the collaboration considered on the one hand an effective operator approach, which is not applicable to our model.
A second part of the analysis considers a s-channel mediator, which couples in a specific way with quarks and the DM particles. In considering the s-channel mediator being given by the $Z$-boson, constraints on the present model parameters can be obtained from the LHC data in a similar analysis.
Such an analysis is beyond the scope of the present paper.

\section{Conclusions \label{conclusion}}
In this paper we considered an $\mathbb{A}_4 \times {\mathbb Z}_2 \times {\mathbb Z}_2^\prime$-symmetric extension of the Standard Model (SM), wherein the $SU(2)_L$-doublet leptons are embedded into $\mathbb{A}_4$-triplet fields and the matter content also includes a ${\mathbb Z}_2$-odd Majorana neutrino.
The scalar sector consists of two $\mathbb{A}_4$-triplet $SU(2)_L$-doublet fields $\Phi^{(k=1,2)}$ among which one is charged under the ${\mathbb Z}_2$ symmetry, and further a SM-singlet $\mathbb{A}_4$-triplet ${\mathbb Z}_2^\prime$-odd scalar field, whose vacuum expectation value (VEV) breaks $\mathbb{A}_4$ and ${\mathbb Z}_2^\prime$.
This renormalisable model has previously been shown to successfully account for the charged lepton and neutrino masses and mixings through a specific alignment of the VEVs.
The small masses of the active neutrinos are generated with a radiative seesaw mechanism. The scale of the $\mathbb{A}_4$ and ${\mathbb Z}_2^\prime$ breaking VEV can be as small as ${\cal O}$(TeV) and thus remain comparatively close to the electroweak scale. We presented the $\mathbb{A}_4 \times {\mathbb Z}_2 \times {\mathbb Z}_2^\prime$-invariant scalar potential for the considered field content and discussed under which circumstances the specific VEV alignment -- which gives rise to large leptonic mixings -- can be realised in this model.

An interesting feature of the model is that the second and third generation  ${\mathbb Z}_2$-even scalars have couplings to leptons that are exclusively lepton flavour violating (while the first generation of the Higgs-like, ${\mathbb Z}_2$-even scalar fields, has coupling strength to leptons that is very close to the corresponding coupling of the SM-Higgs boson).
We have discussed phenomenological constraints on the masses and mass-splittings for the extra ${\mathbb Z}_2$-even scalar fields from the oblique parameters and direct searches for lepton flavour violation at low energy and at LEP. 
The strongest lower bound on the mass scale is 140 GeV due to the constraints from the MEG collaboration on $\mu \to e \gamma$.
At the LHC, this model should have created around ${\cal O}(10)$ lepton flavour violating events in the current data such as $\mu \tau j j$ and $e \tau j j$, where $j$ is an energetic jet, and less than ${\cal O}(1)$ events of $e \mu j j$ events. 
It is possible to interpret the excess in $\mu \tau$ events found by the CMS collaboration as being caused by the fields $H_2,\,\tilde H_2$, which implies that the MEG collaboration should observe $\mu \to e \gamma$ very soon.
Alternatively, a perturbation of ${\cal O}(0.1)$ of the VEV alignment can lead to a LFV branching ratio of the SM-like Higgs boson $H_1 \to \mu \tau$ that can explain the observed excess at the LHC.

The ${\mathbb Z}_2$-odd field content of the model yields a ``dark sector'' with the DM candidates being the Majorana neutrino and the neutral components of the $\mathbb{A}_4$-triplet $SU(2)_L$-doublet $\Phi^{(1)}$, among which we investigated the properties of the scalar DM candidate. 
First we discussed the existing limits from LEP I and II. In particular, the OPAL collaboration provides constraints on the masses and mass splittings for scalar DM with masses below 100 GeV, which turn out to be negligible when considering the DM properties of the model.
Then we derived the constraints on the model parameter space from the observed dark matter relic density and from direct search experiments. 
The experimental constraints from the LUX collaboration imply that the DM particle barely interacts with the Higgs boson when its mass is ${\cal O}(100)$ GeV. 

We found that it is possible to account for the observed relic density for DM with a mass in the interval between 47 and 74 GeV or in the interval 600 GeV and 3.6 TeV.
An important parameter in the model is the mass-splitting between the neutral DM particle and its charged ``sibling'', and also the mass-splitting between the three generations of DM particles. 

It is promising to establish the connection between Dark Matter and the scalars at the LHC and especially at future lepton colliders by studying the decay properties of the $125.7$ GeV SM-Higgs-like scalar with higher precision. 
Last but not least, direct searches in the high-energy data from the next run of the LHC and at future hadron colliders might reveal the extended scalar field content of this model by detecting di-lepton excesses, or multiple (scalar) resonances for centre-of-mass energies of several TeV.

\section*{Acknowledgements}

This project has received funding from the Swiss National Science Foundation.
This project has received funding from the European Union's Seventh Framework Programme for research, technological development and demonstration under grant agreement no PIEF-GA-2012-327195 SIFT.
IdMV thanks the University of Basel for hospitality and Antonio Cárcamo Hernández for useful discussions.

\appendix

\section{The product rules for $\mathbb{A}_4$ \label{A}}

The following product rules for the $A_{4}$ group were used in the construction of the Lagrangian:
\begin{eqnarray}\label{prod-rule-1}
&& \hspace{18mm }\mathbf{3}\otimes \mathbf{3}=\mathbf{3}_{s}\oplus \mathbf{3}_{a}\oplus
\mathbf{1}\oplus \mathbf{1}^{\prime }\oplus \mathbf{1}^{\prime \prime },\\[3mm]
\label{A4-singlet-multiplication}
&&\mathbf{1}\otimes \mathbf{1}=\mathbf{1},\hspace{5mm}\mathbf{1}^{\prime}\otimes \mathbf{1}^{\prime \prime }=\mathbf{1},\hspace{5mm}
\mathbf{1}^{\prime }\otimes \mathbf{1}^{\prime }=\mathbf{1}^{\prime \prime },
\hspace{5mm}\mathbf{1}^{\prime \prime }\otimes \mathbf{1}^{\prime \prime }=\mathbf{1}^{\prime },
\end{eqnarray}
Denoting $\left( x_{1},y_{1},z_{1}\right) $ and
$\left(x_{2},y_{2},z_{2}\right) $ as the basis vectors for two  $A_{4}$-triplets $\mathbf{3}$, one finds:
\begin{equation}\label{triplet-vectors}
\begin{array}{lcl}
\left( \mathbf{3}\otimes \mathbf{3}\right)_{\mathbf{1}}&=&x_{1}y_{1}+x_{2}y_{2}+x_{3}y_{3},\\
\left( \mathbf{3}\otimes \mathbf{3}\right)_{\mathbf{1}^{\prime}}&=&x_{1}y_{1}+\omega x_{2}y_{2}+\omega ^{2}x_{3}y_{3},\\
\left( \mathbf{3}\otimes \mathbf{3}\right)_{\mathbf{1}^{\prime\prime }}&=&x_{1}y_{1}+\omega ^{2}x_{2}y_{2}+\omega x_{3}y_{3},\\
\left( \mathbf{3}\otimes \mathbf{3}\right)_{\mathbf{3}_{s}}&=&\left(x_{2}y_{3}+x_{3}y_{2},x_{3}y_{1}+x_{1}y_{3},x_{1}y_{2}+x_{2}y_{1}\right),\\
\left( \mathbf{3}\otimes \mathbf{3}\right)_{\mathbf{3}_{a}}&=&\left(x_{2}y_{3}-x_{3}y_{2},x_{3}y_{1}-x_{1}y_{3},x_{1}y_{2}-x_{2}y_{1}\right),
\end{array}
\end{equation}
where $\omega =e^{\ci 2 \pi/3}$. The representation $\mathbf{1}$
is trivial, while the non-trivial $\mathbf{1}^{\prime}$ and $\mathbf{1}^{\prime \prime}$
are complex conjugate to each other.
Comprehensive reviews of discrete symmetries in particle physics can be found in refs. \cite{Altarelli:2010gt,Ishimori:2010au, King:2013eh}.

\def\one{\mathbf{1}}
\def\oneP{{\mathbf{1}^{\prime}}}
\def\onePP{{\mathbf{1}^{\prime\prime}}}
\def\threeS{{\mathbf{3}_s}}

\section{Scalar potentials with $\mathbb{A}_4$ triplets \label{B}}

The most general potential for two fields, $\Phi^{(1)}$, $\Phi^{(2)}$ charged under separate $\mathbb{Z}_2$ and simultaneously $\mathbb{A}_4$ triplets and $SU(2)$ doublets
has a large number of possible contractions:
\begin{align*}
    V_{\phi\phi} =& \sum_{i=1,2} \left[
\mu_i^2 \,
\contraction{(}{\cfield}{}{\field}
(\cfield\field)_\one
+
\lambda_{1,(i)} \,
\contraction{(}{\cfield}{}{\field}
\contraction{(\cfield\field)_\one(}{\cfield}{}{\field}
(\cfield\field)_\one(\cfield\field)_\one
+
\lambda_{2,(i)} \,
\contraction{(}{\cfield}{\cfield)_\one(}{\field}
\contraction[1.5ex]{(\cfield}{\cfield}{)_\one(\field}{\field}
(\cfield\cfield)_\one(\field\field)_\one
\right .\\
&+ \left.
\lambda_{3,(i)} \,
\contraction{(}{\cfield}{\field)_\one(}{\field}
\contraction[1.5ex]{(\cfield}{\field}{)_\one(\field}{\cfield}
(\cfield\field)_\one(\field\cfield)_\one
+
\lambda_{4,(i)} \,
\contraction{(}{\cfield}{}{\field}
\contraction{(\cfield\field)_\oneP(}{\cfield}{}{\field}
(\cfield\field)_\oneP(\cfield\field)_\onePP
\right .\\
&+ \left.
\lambda_{5,(i)} \,
\contraction{(}{\cfield}{\cfield)_\oneP(}{\field}
\contraction[1.5ex]{(\cfield}{\cfield}{)_\oneP(\field}{\field}
(\cfield\cfield)_\oneP(\field\field)_\onePP
    \right] \\
&+ 
\kappa_1 \,
\contraction{(}{\cfieldOne}{}{\fieldOne}
\contraction{(\cfieldOne\fieldOne)_\one(}{\cfieldTwo}{}{\fieldTwo}
(\cfieldOne\fieldOne)_\one(\cfieldTwo\fieldTwo)_\one
+
\kappa_2 \,
\contraction{(}{\cfieldOne}{\cfieldTwo)_\one(}{\fieldOne}
\contraction[1.5ex]{(\cfieldOne}{\cfieldTwo}{)_\one(\fieldOne}{\fieldTwo}
(\cfieldOne\cfieldTwo)_\one(\fieldOne\fieldTwo)_\one
+
\kappa_3 \,
\contraction{(}{\cfieldOne}{\fieldTwo)_\one(}{\fieldOne}
\contraction[1.5ex]{(\cfieldOne}{\fieldTwo}{)_\one(\fieldOne}{\cfieldTwo}
(\cfieldOne\fieldTwo)_\one(\fieldOne\cfieldTwo)_\one \\&
+
\kappa_4 \,
\contraction{(}{\cfieldOne}{\fieldOne)_\one(}{\cfieldTwo}
\contraction[1.5ex]{(\cfieldOne}{\fieldOne}{)_\one(\cfieldTwo}{\fieldTwo}
(\cfieldOne\fieldOne)_\one(\cfieldTwo\fieldTwo)_\one
+
\kappa_5 \,
\contraction{(}{\cfieldOne}{}{\cfieldTwo}
\contraction{(\cfieldOne\cfieldTwo)_\one(}{\fieldOne}{}{\fieldTwo}
(\cfieldOne\cfieldTwo)_\one(\fieldOne\fieldTwo)_\one
+
\kappa_6 \,
\contraction{(}{\cfieldOne}{\fieldTwo)_\one(\fieldOne}{\cfieldTwo}
\contraction[1.5ex]{(\cfieldOne}{\fieldTwo}{)_\one(}{\fieldOne}
(\cfieldOne\fieldTwo)_\one(\fieldOne\cfieldTwo)_\one \\&
+
|\kappa_7| \, e^{\ci \beta_7} \,
\contraction{(}{\cfieldOne}{}{\fieldOne}
\contraction{(\cfieldOne\fieldOne)_\oneP(}{\cfieldTwo}{}{\fieldTwo}
(\cfieldOne\fieldOne)_\oneP(\cfieldTwo\fieldTwo)_\onePP
+
\kappa_8 \,
\contraction{(}{\cfieldOne}{\cfieldTwo)_\oneP(}{\fieldOne}
\contraction[1.5ex]{(\cfieldOne}{\cfieldTwo}{)_\oneP(\fieldOne}{\fieldTwo}
(\cfieldOne\cfieldTwo)_\oneP(\fieldOne\fieldTwo)_\onePP
\\&
+
\kappa_9 \,
\contraction{(}{\cfieldOne}{\fieldTwo)_\oneP(}{\fieldOne}
\contraction[1.5ex]{(\cfieldOne}{\fieldTwo}{)_\oneP(\fieldOne}{\cfieldTwo}
(\cfieldOne\fieldTwo)_\oneP(\fieldOne\cfieldTwo)_\onePP
+
|\kappa_{10}| \, e^{\ci \beta_{10}} \,
\contraction{(}{\cfieldOne}{\fieldOne)_\oneP(}{\cfieldTwo}
\contraction[1.5ex]{(\cfieldOne}{\fieldOne}{)_\oneP(\cfieldTwo}{\fieldTwo}
(\cfieldOne\fieldOne)_\oneP(\cfieldTwo\fieldTwo)_\onePP
\\&
+
\kappa_{11} \,
\contraction{(}{\cfieldOne}{}{\cfieldTwo}
\contraction{(\cfieldOne\cfieldTwo)_\oneP(}{\fieldOne}{}{\fieldTwo}
(\cfieldOne\cfieldTwo)_\oneP(\fieldOne\fieldTwo)_\onePP
+
\kappa_{12} \,
\contraction{(}{\cfieldOne}{\fieldTwo)_\oneP(\fieldOne}{\cfieldTwo}
\contraction[1.5ex]{(\cfieldOne}{\fieldTwo}{)_\oneP(}{\fieldOne}
(\cfieldOne\fieldTwo)_\oneP(\fieldOne\cfieldTwo)_\onePP
\\&
+
|\kappa_{13}| \, e^{\ci \beta_{13}} \,
\contraction{(}{\cfieldOne}{}{\fieldTwo}
\contraction{(\cfieldOne\fieldTwo)_\one(}{\cfieldOne}{}{\fieldTwo}
(\cfieldOne\fieldTwo)_\one(\cfieldOne\fieldTwo)_\one
+
|\kappa_{14}| \, e^{\ci \beta_{14}} \,
\contraction{(}{\cfieldOne}{\cfieldOne)_\one(}{\fieldTwo}
\contraction[1.5ex]{(\cfieldOne}{\cfieldOne}{)_\one(\fieldTwo}{\fieldTwo}
(\cfieldOne\cfieldOne)_\one(\fieldTwo\fieldTwo)_\one
\\&
+
|\kappa_{15}| \, e^{\ci \beta_{15}} \,
\contraction{(}{\cfieldOne}{\fieldTwo)_\one(}{\fieldTwo}
\contraction[1.5ex]{(\cfieldOne}{\fieldTwo}{)_\one(\fieldTwo}{\cfieldOne}
(\cfieldOne\fieldTwo)_\one(\fieldTwo\cfieldOne)_\one 
+
|\kappa_{16}| \, e^{\ci \beta_{16}} \,
\contraction{(}{\cfieldOne}{}{\fieldTwo}
\contraction{(\cfieldOne\fieldTwo)_\oneP(}{\cfieldOne}{}{\fieldTwo}
(\cfieldOne\fieldTwo)_\oneP(\cfieldOne\fieldTwo)_\onePP
\\&
+
|\kappa_{17}| \, e^{\ci \beta_{17}} \,
\contraction{(}{\cfieldOne}{\cfieldOne)_\oneP(}{\fieldTwo}
\contraction[1.5ex]{(\cfieldOne}{\cfieldOne}{)_\oneP(\fieldTwo}{\fieldTwo}
(\cfieldOne\cfieldOne)_\oneP(\fieldTwo\fieldTwo)_\onePP
+ \text{h.c.}
\,,
\label{ap:Vphiphi}
\end{align*}
where $\tilde\Phi^{(i)} = \ci \sigma_2 (\Phi^{(i)})^*$ is the charge conjugate of $\Phi^{(i)}$, lines above fields signify $SU(2)$ contractions using the tensor $\epsilon = \ci \sigma_2$ and subscripts of brackets denote the type of $\mathbb{A}_4$ contraction used for the fields within. All couplings are either real or their phase is stated explicitly. All additional $SU(2)$ and $\mathbb{A}_4$ invariant terms can be expressed as linear combinations of the above. Assuming manifest CP symmetry \cite{Holthausen:2012dk}, all couplings must have vanishing imaginary parts.

Likewise if we introduce a real scalar field $\chi$, which is a singlet under the SM gauge group and a triplet under $\mathbb{A}_4$, we find
\begin{equation}
\begin{split}
V_{\chi\phi} =& \sum_{i=1,2} \left[
\sigma_{1,(i)} \,
\contraction{(}{\cfield}{}{\field}
(\cfield\field)_\one(\chi\chi)_\one
+
\sigma_{2,(i)} \,
\contraction{(}{\cfield}{\chi)_\one(}{\field}
(\cfield\chi)_\one(\field\chi)_\one 
\right.\\& \left.
+
|\sigma_{3,(i)}| \, e^{\ci \alpha_{3}^{(i)}} \,
\contraction{(}{\cfield}{}{\field}
(\cfield\field)_\oneP(\chi\chi)_\onePP
+
\sigma_{4,(i)} \,
\contraction{(}{\cfield}{\chi)_\oneP(}{\field}
(\cfield\chi)_\oneP(\field\chi)_\onePP 
\right]
+ \text{h.c.}
\,,
\label{ap:Vchiphi}
\end{split}
\end{equation}
for the potential mixing both sectors and 
\begin{equation}
\begin{split}
V_{\chi\chi} =& 
\,\mu_\chi^2 \,
(\chi\chi)_\one
+
d_1 \,
(\chi\chi)_\threeS(\chi\chi)_\threeS
+
d_2 \,
(\chi\chi)_\oneP(\chi\chi)_\onePP
\end{split}
\label{ap:Vchichi}
\end{equation}
for the potential only depending on $\chi$. In order to make the vacuum as shown in eq.~\eqref{eq:VEVs} an extremal point of the potential, the couplings $\sigma_{3,(2)}$ and $\sigma_{4,(2)}$ must be absent or negligibly small\footnote{For details on the effects when this is not the case, see Appendix \ref{Ap:misalign}.} The mass terms for $\fieldTwo$ and $\chi$ fulfil the relations
\begin{align}
\mu_2^2 &=  -(\lambda_{1,(2)} + \lambda_{2,(2)} + \lambda _{3,(2)}) v^2 -\sigma_{1,(2)} v_\chi^2 \,,\\
\mu_{\chi }^2 &= -\frac{1}{2} \left((4 d_1 + d_2) v_\chi^2 + \sigma_{1,(2)} v^2 \right) \;.
\end{align}

\section{Solution to the pathological vacuum \label{Ap:chis}}
\label{AppendixC}
In the sub-potential $V_{\chi\chi}$ while all derivatives at the position $\chi = v_\chi (1,0,-1)$ can be set to zero with an appropriate condition for $v_\chi$ or $\mu_\chi^2$, the mass square eigenvalues for the $\chi$ subsector are then not positive definite. 

This is an issue for the model being considered as it relies on the VEV $\chi = v_\chi (1,0,-1)$, namely for obtaining eq.~\eqref{Mnu}. The problem can be solved for instance by allowing $r \sim 1$ or by a redefinition of the VEV direction in $\mathbb{A}_4$ space, to either $(1,1,1)$, a permutation of $(1,0,0)$ or even a complex direction $(1,0,i)$. In those cases however, we lose the connection to the flavour model in \cite{Hernandez:2013dta}, and it remains to be shown that large leptonic mixing angles are still possible. Due to the vanishing entry, the latter VEV direction $(1,0,i)$ has comparable phenomenology to $(1,0,-1)$ in terms of leptonic mixing angles (cf.\ eq.~(\ref{Mnu})), but a detailed analysis of this is beyond the scope of the present paper. We note that the change of the VEV direction in $\mathbb{A}_4$ space does not affect the DM analysis in section \ref{pheno}, due to the effective parametrisation used. However, changing the VEV has a strong effect on everything else.

The solution we prefer to use here is to introduce another complex singlet scalar field $\xi$ in the ${\bf 1''}$ representation of $\mathbb{A}_4$. Its interactions to $\chi$ lead to added terms in the scalar potential
\begin{equation}\label{eq:Vxi}
    V_{\xi} = -\mu_\xi^2 \, \xi^\dagger \xi 
        - \gamma_1 \, \xi (\chi \chi)_\oneP
        + \gamma_2 \, \xi \xi \xi
        + \lambda_\xi \, \xi^\dagger \xi^\dagger \xi \xi
        + \lambda'_\xi \, \xi \xi (\chi \chi)_\onePP 
        + \text{h.c.} \;.
\end{equation}
Together with the terms in $V_{\chi\chi}$, it can be shown that this potential can lead to (depending on the parameters) a VEV $\langle \xi \rangle = v_\xi \omega^2$ and the aforementioned $(1,0,-1)$ VEV direction for $\chi$, with positive mass squares of similar magnitude for all mass eigenstates around this vacuum. Since the terms $\xi$ and $(\chi\chi)_\onePP$ have the same assignment under all symmetries, the presence of $\xi$ also leads to additional couplings of form similar to the $\sigma_{3,(i)}$ term of $V_{\chi\phi}$. However, due to the form of $\langle \chi \rangle$ one finds that
\begin{equation}
    \langle (\chi\chi)_\onePP \rangle = -\frac{v_\chi^2}{v_\xi} \, \langle \xi \rangle \,, 
    \qquad
    \langle (\chi\chi)_\oneP \rangle = -\frac{v_\chi^2}{v_\xi^2} \, \langle \xi \rangle^2 \,,
\end{equation}
which means that these additional couplings between $\Phi^{(i)}$ and $\xi$ can simply be absorbed into $\sigma_{3,(i)}$ as far as masses for the $\Phi^{(i)}$ and the VEV alignment (with potential pertubations) are concerned. 


\end{document}